\documentclass[aps,amsmath,twocolumn,amssymb,floatfixng,showpacs,superscriptaddress,footinbib]{revtex4-1}
\usepackage[english]{babel}
\usepackage{amsmath}
\usepackage{amssymb}
\usepackage{natbib}
\usepackage{graphicx}
\usepackage{bm}
\usepackage{soul}
\usepackage[normalem]{ulem}

\usepackage{subfigure}
\usepackage{float}
\usepackage{epsfig}
\usepackage{hyperref}
\hypersetup{
colorlinks=true,final=true,
        linkcolor=blue,
        citecolor=blue,
        filecolor=blue,
        urlcolor=blue,
}

\def\be{\begin{equation}}
\def\ee{\end{equation}}
\def\ba{\begin{eqnarray}}
\def\ea{\end{eqnarray}}

\newcommand{\angstrom}{\text{\normalfont\AA}}

\begin{document}

\title{Third-order Hall effect in the surface states of a topological insulator 
}
\author{Tanay Nag} 
\affiliation{Department of Physics and Astronomy, Uppsala University, Box 516, 75120 Uppsala, Sweden}
\author{Sanjib Kumar Das}
\affiliation{Department of Physics, Lehigh University, Bethlehem, Pennsylvania, 18015, USA}
\affiliation{Leibniz Institute for Solid State and Materials Research (IFW) Dresden, Helmholtzstrasse 20, 01069 Dresden, Germany}

\author{Chuanchang Zeng}
\email{czeng@bit.edu.cn}
\affiliation{Centre for Quantum Physics, Key Laboratory of Advanced Optoelectronic Quantum Architecture and Measurement(MOE),
School of Physics, Beijing Institute of Technology, Beijing, 100081, China}
\affiliation{Beijing Key Lab of Nanophotonics $\&$ Ultrafine Optoelectronic Systems,
School of Physics, Beijing Institute of Technology, Beijing, 100081, China}

\author{Snehasish Nandy}
\email{snehasish12@lanl.gov}
\affiliation{Theoretical Division, Los Alamos National Laboratory, Los Alamos, New Mexico 87545, USA}
\affiliation{Center for Nonlinear Studies, Los Alamos National Laboratory, Los Alamos, New Mexico 87545, USA}

\begin{abstract} 
Time reversal and inversion symmetric materials fail to yield linear and nonlinear responses since they possess net zero Berry curvature. However, higher-order Hall response can be generated in these systems upon constraining the crystalline symmetries. Motivated by the recently discovered third-order Hall (TOH) response mediated by Berry connection polarizability, namely, the variation the Berry connection with respect to an applied electric field, here we investigate the existence of such Hall effect in the surface states of hexagonal warped topological insulator (e.g., Bi$_2$Te$_3$)
under the application of electric field only.
Using the semiclassical Boltzmann formalism, we investigate the effect of tilt and hexagonal warping on the Berry connection polarizability tensor and consequently, the TOH effect provided the Dirac cone remains gapless. We find that the magnitude of the response increases significantly with increasing the tilt strength and warping and therefore, they can provide the tunability of this effect. In addition, we  also explore the effect of chemical doping on TOH response in this system. Interestingly, we show based on the symmetry analysis, that the TOH can be the leading-order response in this system which can directly be verified in experiments.
\end{abstract}

\maketitle

\section{Introduction}
\label{intro}	
In recent times topological transport has gained immense interests since its technological applications are versatile in engineering magnetic and electrical devices \cite{Hasan2010,Hasan2011,Qi2011,Pesin2012,Ando2013}.  
Such topological response
in the linear regime manifests into various forms, namely quantum Hall effect (QHE)~\cite{klitzing80}, quantum spin Hall effect (QSHE)~\cite{kane05}, quantum anomalous Hall effect (QAHE)~\cite{haldane88,liu16}, anomalous Nernst effect~\cite{Xiaonernst}, and so on. The emergence of different kind of responses strongly depends on the interplay of topology and symmetry of the system. For instance, generation of the QHE requires application an external magnetic field, whereas, QSHE takes place in absence of magnetic field demanding the system to have time-reversal symmetry (TRS). On the other hand, QAHE originates from the intrinsic magnetic moment present in the system which also breaks TRS. In essence, it is the geometric nature of the wavefunctions, also called Berry curvature (BC) \cite{Berry1984,Niu_2010}, which is at the root of all such topological responses. While normal Hall like response originates in magnetic systems, it has been shown recently that nonmagnetic systems without inversion symmetry (IS) and  in presence/absence of mirror symmetry  can give rise to higher-order Hall response, namely second-order nonlinear Hall effect (NLHE) \cite{sodemann15,Nandy_21, Nandy2019,Matsyshyn2019,Das21,Nag22a,Sadhukhan21a,Sadhukhan21b}. These responses are  either caused by the BC itself or its first moment i.e., BC  dipole of all the occupied states. Importantly,  the nonlinear response is a Fermi surface dependent quantity coupled with the BC.

It is well-known that systems where both IS and TRS are present, the BC is identically zero in magnitude. Interestingly, it is proposed very recently that such systems can in principle give rise to
third-order Hall (TOH) like response. 
Such response is originated from the variation of field induced Berry connection ($\mathcal{A}^{(1)}$) with respect to an applied electric field ($\mathbf{E}$). Even though, the Berry connection in itself is not a gauge invariant quantity, $G_{ab}(\bm k)=\dfrac{\partial \mathcal{A}^{(1)}_{a}(\bm k)}{\partial E_{b}}$, termed as Berry connection polarizability (BCP) is a gauge invariant quantity, where $a$, $b$ represent Cartesian coordinates~\cite{Liu2021, Wei_2022, Su_2022}. Polarizability in electrodynamics indicates an affinity of a matter to gain electric dipole moment in presence of an applied electric field. One can similarly argue that Bloch electrons acquire positional shift due to the external electric field yielding third-order Hall effect (TOHE) \cite{Gao2015}.

Generically, there are plethora of noncentrosymmetric Dirac and Weyl materials which have been proposed to realize NLHE \cite{Low2015,Zhang2018,BCD_2021,Facio2018,Du2018,You2018,Ma2018,Zhang2018I,Du2019,Konig2019,Kang2019,Son2019,Battilomo2019,Zeng2019,Yu2019,
Kumar2021,Ortix2021,Du2021}. Most of these systems possess large BC centered around the  Dirac or Weyl nodes. Apart from searching NLHE in TRS preserving but IS breaking materials, there still exist a large class of material which have both TRS and IS. For those systems, it is of natural  interest to understand  higher-order Hall response, namely TOHE there. Recently, the experimental works on T$_{d}$-MoTe$_2$ \textcolor{black}{and few-layer WTe$_2$ flakes are} reported to possess such third-order response \cite{Lai2021, Min_2022}. 
In-plane circular photogalvanic effect on 1T$^{'}$-MoTe$_{2}$, caused by third-order nonlinear optical effect, has been experimentally observed recently \cite{Ma2021}.

Motivated by the above experiments, in this work, we
seek the answer for the following question: How does 
TOHE (i.e., $\propto \mathbf{E}^3$) show up on the surface states of a strong three-dimensional (3D) topological insulator (TI)? Using the framework of semiclassical Boltzmann transport theory, we first consider the general expression of TOH current, mediated by BCP,  which is linearly proportional to relaxation time ($\tau$). Our study indicates that the TOH response can appear as the leading-order response in a TRS invariant system containing non-tilted Dirac cone; this can directly be verified in experiments. The magnitude of the response enhances significantly by increasing the hexagonal warping strength which is inherently present in the surface states of 3D TI. This is originated from the fact that BCP tensor acquires higher values with increasing warping.
The TOHE becomes more pronounced with band tilt that breaks $C_{3}$ symmetry, however, the BCP tensor is insensitive with respect to tilt. Moreover, we also explore the effect of chemical doping on TOH response, and discuss in detail how to separate the TOH response from linear and second-order effect in experiments using the symmetry arguments.

The rest of the article is organized as follows. In Sec.~\ref{formalism}, we present the detailed formalism of semiclassical Boltzmann transport theory for TOHE. Following this, in Sec.~\ref{model} we elucidate the TOHE in the surface states of a 3D TI both in the absence and presence of hexagonal warping. Finally, in Sec.~\ref{conc} we conclude with summarizing our results and discussing possible future directions. 


\section{Semiclassical Formalism of TOHE}
\label{formalism}

In this section, we present the general expression of BCP induced TOH conductivity in the absence of external magnetic field within the framework of Boltzmann transport formalism with the relaxation time approximation \cite{Gao2014,Gao2015,Liu2021}. We start with the Boltzmann transport equation with its phenomenological form ~\cite{ashcroft1976solid,ziman2001electrons}
\begin{equation}
\left(\frac{\partial}{\partial t}+\mathbf{\dot{r}}\cdot\mathbf{\nabla_{r}}+{\dot{\bm k}}
\cdot {\nabla_{\bm k}}\right)f_{{\bm k},\mathbf{r},t}=I_{coll} \{f_{{\bm k},\mathbf{r},t}\},
\label{eq_BZ}
\end{equation}
where the right side $I_{coll} \{f_{{\bm k},\mathbf{r},t}\}$ is the known as the collision integral
incorporating the effects of electron correlations and impurity scattering. The non-equilibrium electron distribution function is denoted by  $f_{\mathbf{k},\mathbf{r},t}$. Now under the relaxation time
approximation the steady-state Boltzmann equation reads as
\begin{equation}
(\mathbf{\dot{r}}\cdot\mathbf{\nabla_{r}}+{\dot{\bm k}}\cdot{\nabla_{\bm k}})f_{{\bm k}}=\frac{f_{0}-f_{{\bm k}}}{\tau({\bm k})},
\label{eq_BZf}
\end{equation}
where $\tau(\bm k)$ is the scattering time. For simplicity, we ignore the  momentum dependence of $\tau(\bm k)$ in all the calculations and assume it to be a constant~\cite{Nandy_2017,Nag_2020}. The equilibrium distribution function $f_0({\bm k})$ in absence of applied electric field ${\bm E}$ is given by the Fermi-Dirac distribution function,  
\be 
f_0({\bm k}) = \dfrac{1}{1 + e^{\beta\left[\epsilon({\bm k}) - \mu\right]} }~,
\label{fd_statics}
\ee
where $\beta = 1/(k_B T)$, $\epsilon_ {\bm k}$ and ${\mu}$ are the energy dispersion and chemical potential, respectively.

To study the BC induced linear Hall effect, first-order correction of the band energy due to the orbital magnetic moment is sufficient. This is because of the fact that the orbital magnetic moment couples to the applied magnetic field $\mathbf{B}$, giving rise to an anomalous velocity component  for the electrons~\cite{Niu_2010}. Conversely, systems, preserving TRS, can yield second-order Hall effect under the application of a strong
enough electric field. This occurs by the virtue of the dipole moment of the BC, which in this case generates the anomalous velocity component~\cite{sodemann15}. However, in the case of TOHE, one needs a second-order semiclassical theory for Bloch electrons under uniform electromagnetic fields in terms of physical position and crystal momentum which are fully gauge invariant. 
This theory includes a first(second)-order field correction to the BC (band energy) and modifies the relation between the physical position and crystal momentum with regard to the canonical ones~\cite{Gao2014}. To be precise, being
perturbed by an electric field $\bm{E}$ with $H_E^{\prime}=e\bm{E}\cdot \bm{(r-r}_{c})$, the wavepacket acquires a positional shift with respect to its center $\bm{r}_{c}$ in terms of the second order correction in electric field \cite{Gao2014}.

Now including n$^{th}$-order field corrections to the BC ${\bm \tilde{\Omega}_{\bm k}}$ and band energy $\tilde{\epsilon}_k$, the semiclassical equations of motion in the absence of magnetic field can be written as 
\ba 
\dot {\bm r} &=& \frac{1}{\hbar}\nabla_{\bm k}\tilde{\epsilon}_k
-\dot{\bm{k}} \times {\bm \tilde{\Omega}_{\bm k}},\quad \hbar \dot {\bm k}= e {\bm E}~,
\label{eq_rdot}
\ea
with $e<0$. Here, $\tilde{\epsilon}_k$ and ${\bm \tilde{\Omega}_{\bm k}}$ are  given by
\ba 
&&\tilde{\epsilon}_{\alpha,\bm{k}}=\sum_{i=0}^{n}{\epsilon}_{\alpha,\bm{k}}^{(i)}, \quad {\bm \tilde{\Omega}_{\alpha \delta,\bm k}}=\bm{\nabla_{\bm{k}}} \times \sum_{i=0}^{n}\bm{A}_{\alpha \delta}^{(i)}(\bm{k}),
\label{mod_berry}
\ea
where ${\epsilon}_{\alpha,k}^{(0)}$ and $\bm{A}_{\alpha \delta}^{(0)}(\bm{k})$ are the unperturbed band energy and interband Berry connection or non-Abelian Berry connection matrix, respectively, where $\bm{A}_{\alpha \delta}^{(0)}(\bm{k})$ can be expressed as $\bm{A}_{\alpha \delta}^{(0)}(\bm{k})=\langle u^{(0)}_{\alpha k}|i\bm{\nabla_{\bm{k}}}|u^{(0)}_{\delta k}\rangle$, with $|u^{(0)}_{\alpha k}\rangle$ is the cell-periodic part of the Bloch eigenstate in the unperturbed case  and $\alpha,  \delta$ are band indices. In this work, we restrict ourselves to $n=1$ and $n=2$ for BC and band energy respectively.

Assuming minimal coupling and using standard perturbation theory, the first-order ${\mathcal O}(E)$ correction to
Bloch wavefunction can be written as
\ba 
|u^{(1)}_{\alpha k}\rangle=\sum_{\delta \neq \alpha}\frac{|u^{(0)}_{\delta k}\rangle\langle u^{(0)}_{\delta k}|{H}^{\prime}_E|u^{(0)}_{\alpha k}\rangle}{\epsilon_{\alpha,k}^{(0)}-\epsilon_{\delta,k}^{(0)}}=\sum_{\delta \neq \alpha}\frac{e\bm{E}\cdot \bm{A}^{(0)}_{\delta \alpha}|u^{(0)}_{\delta k}\rangle}{\epsilon_{\alpha,k}^{(0)}-\epsilon_{\delta,k}^{(0)}}, \nonumber \\
\label{en_corr}
\ea
where $\bm{r}=i\partial_{\bm{k}}$ has been applied. \textcolor{black}{However, the first-order correction to the band energy  vanishes i.e., $\epsilon^{(1)}_\alpha=\langle u^{(0)}_{\alpha k}|{H}^{\prime}_E|u^{(0)}_{\alpha k}\rangle=0$} \textcolor{black}{as $\langle u^{(0)}_{\alpha k}|\bm{r}|u^{(0)}_{\alpha k}\rangle=\bm{r}_c$}~\cite{Gao2014}. Importantly, the first-order  Berry connection, measuring  a shift in its center of
mass position  wave packet, incorporates ${\mathcal O}(E)$ written as 
\ba
A_{\alpha,a}^{(1)}&
=&2{\rm Re}\textcolor{black}{\sum_{\delta}^{\delta \neq \alpha}}\frac{A^{(0)}_{\alpha \delta,a} A^{(0)}_{\delta \alpha,b}}{\epsilon_{\alpha,k}^{(0)}-\epsilon_{\delta,k}^{(0)}}E_b=G_{\alpha,ab}E_b.
\label{BC_corr}
\ea
Here, $A_{\alpha,a}^{(1)}$ gives the positional shift for the band $\alpha$ and $G_{ab}$ is known as the BCP tensor and is a purely geometric quantity.
Note that the center of mass position is given by $\mathbf{r}_{c,\alpha} = \mathbf{r}_{0,\alpha} + \mathbf{A}^{(0)}_{\alpha} + \mathbf{A}^{(1)}_{\alpha}$ where $\mathbf{r}_{0,\alpha}$ is a constant, $\mathbf{A}^{(0)}_{\alpha}  \sim O(E^{0})$ is gauge dependent and $\mathbf{A}^{(1)}_{\alpha}\sim O(E^{1})$ involves gauge independent Berry connection polarizability. It is important to note that the
zeroth-order energy can be effectively considered as $\epsilon^{(0)}=\epsilon_k +  \mathbf{E} \cdot \mathbf{r}_c$ such that the total perturbative Hamiltonian
$H_E=\bm {E\cdot r}=\bm { E\cdot (r - r_c) + E\cdot r_c}$. However, this additional term is $k$-independent acting like a potential energy, leading to an overall shift of the energy. On the other hand, in the second-order semiclassical theory, the $r_c$ and the momentum $k$ are viewed as independent variables. Therefore, it is clear $\frac{\partial \epsilon^{(0)}}{\partial k}=\frac{\partial \epsilon_k}{\partial k}$ implying that the additional term will not play any role in our calculation~\cite{xiang2023third}.
Now, the second-order ${\mathcal O}(E^2)$ correction to band energy becomes
\ba
\epsilon^{(2)}_\alpha=\sum_{\delta \neq \alpha}\frac{|\langle u^{(0)}_{\delta k}|{H}_E^\prime|u^{(0)}_{\alpha k}\rangle|^2}{\epsilon_{\alpha,k}^{(0)}-\epsilon_{\delta,k}^{(0)}}=\frac{e^2 E_a G_{ab}E_b}{2}.
\ea
Note that $\epsilon^{(1)}_\alpha$ and $\bm{A}^{(0)}_{\alpha}$ ($\epsilon^{(2)}_\alpha$ and $\bm{A}^{(1)}_{ \alpha}$) are gauge dependent (independent) quantities.

Considering uniform electric field (i.e., $\dot {\bm r} \cdot \nabla_{\bm r}f(\bm k)=0$), the semiclassical Boltzmann equation in Eq.~\eqref{eq_BZf} reads
\ba
\dot{\bm k}\cdot \nabla_{\bm k}f_{\bm{k}} = \tau^{-1}(f_{0}-f_{\bm{k}}).
\label{uniBZ}
\ea
One can consider the following ansatz as the solution of the above Boltzmann equation 
\ba
f_{\bm{k}} = \sum_{m=0}^{\infty}(\tau \bm E \cdot \nabla_{\bm k})^{m} f_{0}(\tilde{\epsilon_{\bm{k}}}).
\label{dis_fun}
\ea
Here we expand the solution in terms of the external electric field which is considered to be small in the semiclassical regime. The ansatz chosen in Eq. (\ref{dis_fun}) reduces to the equilibrium distribution function $f_0 (\epsilon^{(0)}_{\bm k})$ for $m=0$. Note that the ansatz is usually chosen with the band velocity and energy derivative of equilibrium Fermi-function $\partial f_0/\partial \epsilon^{(0)}_{\bm k}$. In the present way of representation, the velocity $\tilde{v}_{\bm k}=\frac{1}{\hbar}\frac{\partial \tilde{\epsilon_{\bm k}} }{\partial {\bm{k}}}$ is replaced with the momentum derivative $\nabla_{\bm k}$ while the Fermi-distribution function contains the  modified energy $f_0 (\tilde{\epsilon_{\bm k}})$ such that effect of the electric field is not double counted. As already discussed, modified energy contains electric field-induced correction terms $\tilde{\epsilon_{\bm k}}= \epsilon^{(0)}_{\bm k} + \epsilon^{(1)}_{\bm k} + \epsilon^{(2)}_{\bm k}+ \cdots$.

Plugging the expression of $\dot {\bm r}$ and $f_{\bm k}$ given in Eq.~(\ref{eq_rdot}) and Eq.~(\ref{dis_fun}) respectively, into the expression of current density $\mathbf{j} = e \int [d\bm k] \mathbf{\dot{r}} f_{\bm k}$, and on simplifying further, one can obtain the third-order current (i.e., $\propto \bm{E}^3$) which can be written as (considering $e=\hbar=1$) \cite{Liu2021}

\begin{widetext}
\begin{equation}
\begin{split}
\mathbf{j}_3 & = - \int [d\bm k](\mathbf{E}\times\Omega^{(0)}_{\bm k})[\epsilon^{(2)}f^{'}_{0}(\bm k)]+\tau  \int [d\bm k] v_{\bm{k}}^{(0)}(\mathbf{E}\cdot \nabla_{\bm k})[\epsilon^{(2)}f^{'}_{0}(\bm k)] +\tau  \int [d\bm k] v_{\bm{k}}^{(2)}(\mathbf{E}\cdot \nabla_{\bm k})f_{0}(\bm k) \\
& -\tau  \int [d\bm k](\mathbf{E}\times\Omega^{(1)}_{\bm k})(\mathbf{E}\cdot \nabla_{\bm k})f_{0}(\bm k)-\tau^2  \int [d\bm k](\mathbf{E}\times\Omega^{(1)}_{\bm k})(\mathbf{E}\cdot \nabla_{\bm k})^{2}f_{0}(\bm k)+\tau^3  \int [d\bm k] v_{\bm{k}}  (\mathbf{E}\cdot \nabla_{\bm k})^3 f_{0}(\bm k),
\end{split}
\label{details_charge}
\end{equation}
\end{widetext}
where $[d\bm k]$ is the notation for $d^d\bm k/(2\pi)^{d}$ with $d$ being the dimension of the system, $f^{'}_{0}(\bm k)=\frac{\partial f_{0}(\bm k) }{\partial \epsilon_{\bm{k}}}$ and $v_{\bm k}^{(i)}=\frac{\partial \epsilon^{(i)}_{\bm k} }{\partial {\bm{k}}}$. It is clear from the above expression that all the terms contain the energy derivative of Fermi-Dirac distribution function and therefore, the Hall effect associated with this current is purely caused by the  Fermi surface. Here, the first term  is independent of the relaxation time and this is caused by the  the combination of anomalous velocity, induced by BC, and second-order field correction of the band energy. The first term hence yields purely intrinsic TOHE. On the other hand, the second and third terms of Eq.~(\ref{details_charge}) are linearly proportional to $\tau$, appearing due to the second-order energy correction in the distribution function and the velocity, respectively. The fourth and fifth term arise mainly due to the anomalous velocity produced by the field-induced BC. The last term, which is proportional to $\tau^3$, is purely semiclassical and emerges due to band velocity. 

Since in the present work we are interested in contribution of TOHE originating from the BCP, we will drop the semiclassical term from now on. However, we would like to point out that one can separate this term from the others by looking at the $\tau$-scaling in experiment~\cite{Du2019}. It is expected that the third-order current will have very small signal as compared to first-order current in experiment. Therefore, in this work, we will consider the TRS invariant system so that BC mediated  linear anomalous Hall effect
vanishes, which otherwise becomes
dominant for TRS broken systems. It is important to note that the purely intrinsic first term $\propto \tau^0$ and anomalous velocity related fourth term $\propto \tau^2$ of Eq.~(\ref{details_charge}) vanish in the  TRS invariant system. Therefore, the third-order current expression has terms only proportional to $\tau$ for a TRS invariant system. It is also important to note that another contribution proportional to $\tau^2$ in non-linear planar Hall effect can appear in TRS broken systems arising from the combination of unperturbed BC and band energy \cite{Steven19,Zhou21,Battilomo21}. In addition to the purely intrinsic term of Eq.~(\ref{details_charge}), the second-order field
dependent BC arising from the second-order field-induced positional shift can genetare additional contribution to TOH for TRS broken system~\cite{xiang2023third}. However, the second-order field-induced Berry connection contributes only to TOH for TRS broken system in contrast to TRS invariant system where first-order-field-induced Berry connection contributes to TOH.

It is instructive to rewrite the
third-order current given by Eq.~(\ref{details_charge}) in component form as
\be
j_{3,a}=\chi_{abcd}E_{b}E_{c}E_{d}
\label{eqcondtens}
\ee
where the third-order conductivity tensor $\chi_{abcd}$ is a fourth rank tensor which can generate both the longitudinal and transverse third-order current response. Now, from the Eq.~(\ref{details_charge}), the third-order conductivity tensor $\chi_{abcd}$  for a TRS invariant system can be written in terms of BCP tensor as \cite{Liu2021}

\be
\begin{split}
\chi^{(3)}_{abcd} & = \tau \int [d\bm k] [\partial_{a}\partial_{b}G_{cd}-\partial_{a}\partial_{d}G_{bc}+\partial_{b}\partial_{d}G_{ac}]f_{0}(\bm k)\\
&  -\dfrac{\tau}{2} \int [d\bm k] v_{a}^{(0)}v_{b}^{(0)}G_{cd}f^{\prime \prime}_{0}(\bm k),
\end{split}
\label{lienar}
\ee

where we use $\partial_{i}\partial_{j}=\partial_{j}\partial_{i}$,  $G_{ij}=-G_{ji}$, $v_i f'_{0}(\bm k)=\partial_{i} f_{0}(\bm k) $ and $\int [d\bm k] \partial_{i}G_{jk} \partial_{l}f_{0}(\bm k) =- \int [d\bm k] \partial_{i} \partial_{l} G_{jk} f_{0}(\bm k)$.
We now wish to separate the conductivity tensor into the components that contribute to the power and dissipationless Hall components. Considering the fact that the current and
electric fields transform as vectors under coordinate changes, we now decompose the $\chi^{(3)}_{abcd}$ tensor into symmetric $\chi^{(3),S}_{abcd}$ and antisymmetric $\chi^{(3),A}_{abcd}$ parts with respect to the first two indices as $\chi^{(3)}_{abcd}=\chi^{(3),S}_{(ab)cd}+\chi^{(3),A}_{(ab)cd}$ \cite{Liu2021}. Note that $\chi^{(3),A}_{(ab)cd}$ represents dissipationless  TOH conductivity tensor. In connection with Eq.~(\ref{eqcondtens}), the BCP induced third-order response is not constrained by TRS and IS. However, crystalline symmetries are very important to observe TOHE. The TOHE vanishes when the 2D system possesses $C_{3v}$, $C_{6v}$, $D_3$, $D_{3h}$, $D_{3d}$, $D_6$ symmetries while $C_3$ and $C_6$ symmetries force the TOH current to be isotropic. Importantly, the mirror symmetry along $j$, $M_j$  constraints $\chi^{(3)}_{ijjj}=\chi^{(3)}_{iiij}=0$ with $i\ne j$. Interestingly, the spin susceptibility of BCP leads to non-linear planar Hall effect, allowed by  $C_n$, $C_{nv}$ and $D_n$ ($n=2,3,4,6$) symmetries, for TRS broken case \cite{huang2022intrinsic}. 

It is important to note that BC induced first-order, second-order and TOH response can appear simultaneously in experiment for a system with broken both the TRS and IS. However, one can easily separate them from each other via frequency lock-in (ac) measurements, specifically, by measuring second-harmonic and third-harmonic Hall resistance~\cite{Lai2021}. In dc measurements, they can also be
distinguished based on the above symmetry analysis. It has been shown that in a TRS invariant 2D system, the presence of single mirror symmetry forces the BCD induced second-order Hall conductivity to be orthogonal to the mirror plane. On the other hand, the TOH conductivity vanishes in the direction orthogonal to the mirror plane. Therefore,  this very fact can isolate second-order Hall and TOH responses in a TRS invariant but IS broken systems where the linear Hall response already vanishes due to presence of TRS.

Based on the above symmetry analysis, we would now draw remarks on real systems where the BCP induced TOH response will be the dominant one. Unlike the first- and second-order Hall effects, the third-order response is not restricted by TRS and IS. The TOHE appear as the leading one in non-magnetic centrosymmetric materials where both first- and second-order responses are forced to vanish. Apart from the non-magnetic centrosymmetric systems, it is also possible to have the TOHE as the leading one in non-centrosymmetric systems. For example, in 2D, the BCD induced second-order response can be suppressed due to the presence of a two-fold (screw) rotation along $z$ axis and therefore, the third-order response will be the dominant one in this case. The materials, having $C_{3v}$ point group  symmetries, can also show TOHE as the dominant one despite the absence of inversion center.


\section{Results}
\label{model}
In this section, we will investigate the BCP induced Hall conductivity in two-dimensional surface Hamiltonian of a 3D topological system, specifically, in the surface states of a 3D strong TI. 

We consider the two-dimensional surface Hamiltonian of a TRS invariant strong TI, (e.g., Bi$_2$Te$_3$) hosting a unique Fermi surface that encloses an odd number of Dirac cones in the surface Brillouin zone. In this system, the linear $k$-dependent spin-orbit coupling leads to the band inversion  at $\Gamma$ point in the Brillouin
zone. In addition, it contains hexagonal warping term ($\propto k^3$) which can be understood as a counterpart of cubic Dresselhaus spin-orbit coupling. 

The reasons for choosing this system to study TOHE are the following: (i) The linear AHE vanishes due to the presence of TR symmetry, (ii) the BCD induced second-order Hall response is zero in the absence of tilt parameter due to crystalline symmetry (iii) this system allows us to investigate the non-trivial effects of hexagonal warping on TOHE, (iv) our predicted results on TOHE can directly be checked in experiments.

\begin{figure}[htb]
\begin{center}
\epsfig{file=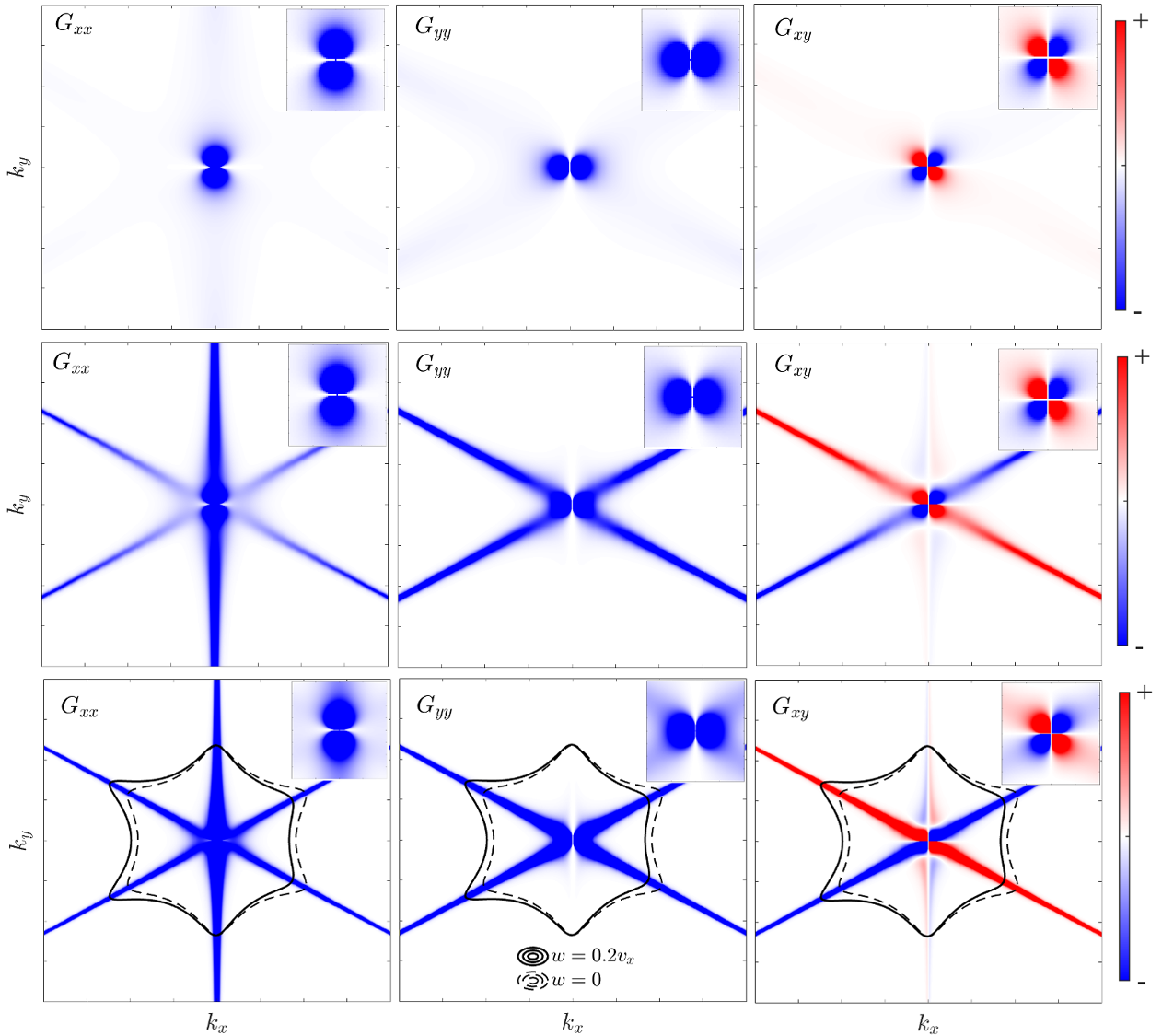,trim=0.0in 0.03in 0.0in 0.03in,clip=true, width = 85mm}\vspace{0em}
\caption{(Color online) Plots show  BCP tensor $G_{xx}$ (left column), $G_{yy}$ (middle column), and $G_{xy}$ (right column)  for  Dirac cone, derived from Eq.~(\ref{ham_hexagonal}),  with warping strength $\lambda=0.05\lambda_0, 0.40 \lambda_0$, and $1.0\lambda_0$ in  top, middle and  bottom row, respectively. With increasing warping strength,  $G_{xx}$ displays snow-flake like structure whereas for $G_{yy}$ and $G_{xy}$ show quadrupolar features for ${\bm k}$ being significantly away from $\Gamma$-point. The four legs of the quadrupole
acquire same sign for $G_{yy}$ while sign changes between two consecutive legs for $G_{xy}$.
Insets show the close view of BCP components near $\Gamma$-point where the diagonal components show  dipole-like structure whereas the off-diagonal component exhibits  quadrupole-like structure. The tilt parameter $w$ does not affect the BCP tensor, it results in a change in the Fermi surface as 
indicated by the black lines in the bottom row for $E_f =0.3~eV$. We consider parameters, representing Bi$_2$Te$_3$, as follows $v_x =v_y =2.55~eV\cdot \angstrom, \lambda_0 =250~eV\cdot \angstrom^3$, the band tilt strengths $w$ are as given in the panels. The color scale used for the insets is 10 times of the main figure's color scale.}
\label{BCP_Tensor}
\end{center}
\end{figure}

Considering the threefold rotation $C_3$ around the $z$ axis and mirror symmetry $M_x$: $x\to -x$, the low-energy model around the gapless $\Gamma$ point is given by 
\begin{equation}
 H^{\rm HW}({\bm k})= E_0 ({\bm k}) + v_x k_x \sigma_y -v_y k_y \sigma_x + \frac{\lambda}{2}(k_+^3 + k_-^3)\sigma_z,
 \label{ham_hexagonal}
\end{equation}
with $E_0(k)= \frac{k^2}{2m^*}$ causes the particle-hole asymmetry which for simplicity is ignored here. Here, $v_x$ and $v_y$ are the Dirac velocities along $x$ and $y$ directions respectively which we consider to be $k$ independent without loss of generality, $k_{\pm}=k_x \pm i k_y$ and $\lambda$ is the strength of hexagonal warping. 


Now the energy dispersion of the above Hamiltonian becomes
\begin{equation}
E^{\pm}({\bm k})= \pm \sqrt{v^2k^2 +  \lambda^2 k^6 \cos\phi},
\label{en_TI}
\end{equation}
where $\phi={\rm arctan}(\frac{k_y}{k_x})$ and $+ (-)$ represents conduction (valence) band. The
band dispersion has sixfold symmetry under $\phi \rightarrow \phi + \frac{2\pi}{6}$. In the absence of warping, it is clear from the Eq.~(\ref{en_TI}) that the Fermi surface, obtained from $f'_{0}(\bm k)$, is circular. After turning on the hexagonal warping term, the Fermi surface remains circular for small strength of warping. With increasing warping strength, the shape of the Fermi surface becomes non-circular with relatively sharp tips extending along a high symmetry direction and curves inward in between, leading to a snowflake-like structure~\cite{Fu09}. Note that, the surface states depicted by the Hamiltonian given in Eq.~(\ref{ham_hexagonal}) preserve TR symmetry.

Interestingly, in-plane 
surface magnetic field, realized by in-plane magnetization doping or the proximity effect of ferromagnetic insulators with in-plane magnetization,  does not gap out the surface Dirac cones.  
However, the position of the Dirac points in the Brillouin zone changes under such in-plane magnetic field causing the anisotropic spin texture \cite{tilt_TSS_2012_zhang, breakingTRS_Zhang_2013PRL}.  Such anisotropy can be effectively considered through a tilt term under certain conditions. In addition, the  electric field can also lead to surface inversion symmetry as well as particle-hole symmetry breaking which can be modeled by the additional tilt term where the effect of the in-plane magnetic field can also be absorbed \cite{somroob2021tunable}. The arbitrary termination of TI can also lead to the breaking of particle-hole symmetry in the surface states \cite{tilt_TSS_2012_zhang, breakingTRS_Zhang_2013PRL}.

Under such a condition, we can
consider a generic surface Hamiltonian in presence of a tilt term $\omega k_x \sigma_0$.  Here $\omega$ is the tilt strength along the $k_x$-direction. In addition, the $C_{3}$ symmetry breaking can naturally bring  perturbation terms to the Hamiltonian in Eq.~(\ref{ham_hexagonal}). Among these the leading contribution can be effectively described by the above tilted term. 
Such a band tilt term has recently been shown to play an important role in the studies of transport phenomena~\cite{PHE_tilt_2020, super_transport_2021,NLNE_czeng_2022}. In this work, we discuss its effects on the TOHE. It is important to note that our analysis is also applicable for the crystalline topological insulator where there exists an effective TRS symmetry by which one tilted Dirac cone gets mapped to other tilted cone under the TRS operation.

\begin{figure}[htb]
\begin{center}
\epsfig{file=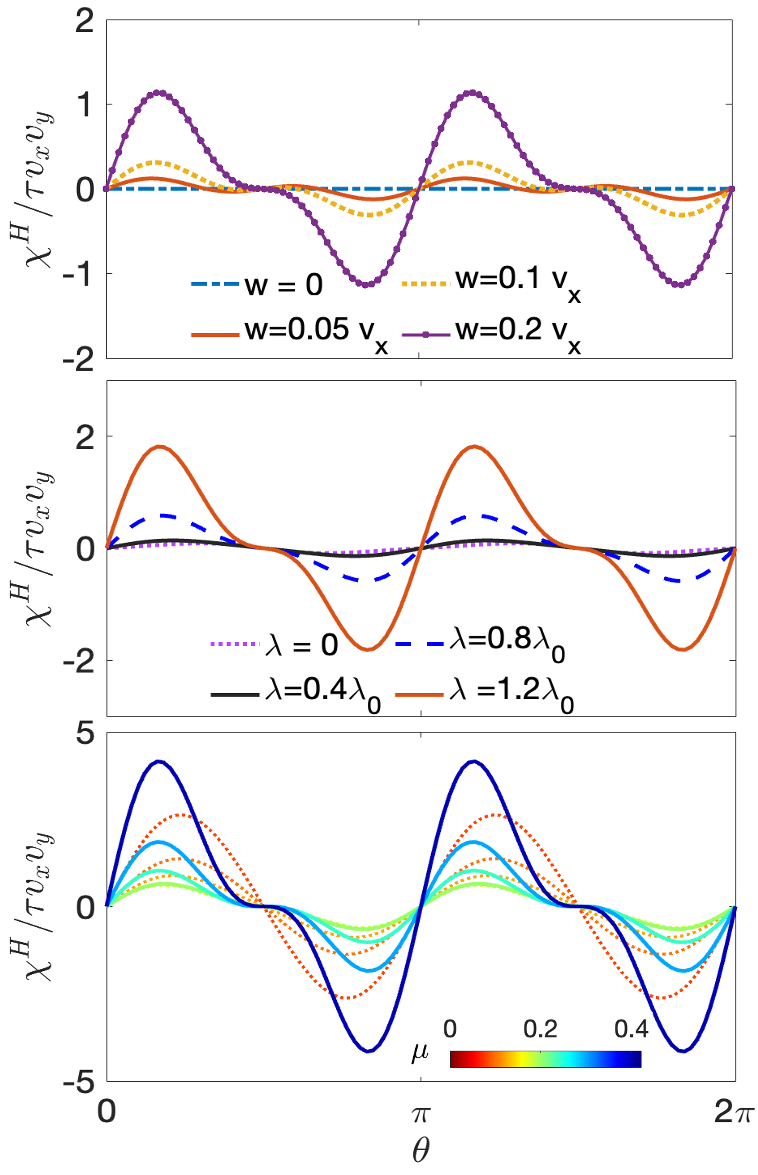,trim=0.0in 0.08in 0.0in 0.05in,clip=true, width=75mm}\vspace{-1em}
\caption{(Color online) Plots depict TOH conductivity (scaled with $\tau v_x v_y$) for the surface states of 3D TI as a function of  the angle $\theta$ between applied electric field and $x$-axis. The angular variation of TOH conductivity with tilt strength $\omega$  (top row) for fixed $\mu=0.25$ eV and $\lambda_0=250$ eV-$\AA^{3}$, warping strength (middle panel) $\lambda$ for fixed $\mu=0.25$ eV and $\omega/v_x=0.2$,  and chemical potential (lower panel) $\mu$ for fixed $\lambda_0=250$ eV-$\AA^{3}$ and $\omega /v_x=0.2$ are shown. The TOH conductivity scaled with $\tau v_x v_y$ is measured in the unit of eV$^{-3}$.}
\label{TOHE_Tensor}
\end{center}
\end{figure}

In order to calculate the third-order conductivity, we first compute the different components of the BCP tensor. One can analytically find the following component
\begin{equation}
    \begin{split}
        G_{xx}=  &~4\big[k_y^2 v_x^2 v_y^2 + 4 k_x^6 v_x^2 \lambda^2 + 9 k_y^2 v_y^2 \lambda^2 (-k_x^2 + k_y^2)^2 \big]/d^3, \\
        G_{yy}=  & ~4 k_x^2 \big[ (k_x^2 + 3 k_y^2)^2 v_y^2 \lambda^2 + v_x^2 (v_y^2 + 36 k_x^2 k_y^2 \lambda^2 )\big]/d^3,\\
        G_{yx} =& -4 \big[ 3k_x^5 k_y (4 v_x^2 + v_y^2) \lambda^2 +6 k_x^3 k_y^3 v_y^2 \lambda^2  \\
        &  ~~~~~~~~~~ +k_x k_y v_x^2 v_y^2 - 9 k_x k_y^5 v_y^2 \lambda^2  \big]/d^3,
    \end{split}
\end{equation}
where $d=(k_x^2 v_x^2 + k_y^2 v_y^2 + k_x^2 \lambda^2 (k_x^2-3 k_y^2)^2)^{1/2}$.
The distribution of $xx$, $yy$ and $xy$ components  are depicted in Fig.~\ref{BCP_Tensor} upper, middle and lower panel for warping strength $\lambda=0.05\lambda_0, 0.40 \lambda_0$, and $1.0\lambda_0$, respectively. The Fermi surface of the system for warping strength $\lambda=250~eV\cdot\AA^{3}$ and band tilt $w =0, 0.2 v_x$ is shown in Fig.~\ref{BCP_Tensor}
lower panel. 
It is clear from the insets in Fig.~\ref{BCP_Tensor} that near the $\Gamma$ point, the diagonal components of the BCP show dipole-like structure (dipole along $y$ [$x$] for $G_{xx}$ [$G_{yy}$]) whereas the off-diagonal component (G$_{xy}$) exhibits a quadrupole-like structure for a particular strength of $\lambda$. Interestingly, although the diagonal components do not change sign, the off-diagonal components shows sign change with $\pi$-periodicity.
This can be well understood from the approximated analytical form of BCP components for $\lambda \to 0$,
 \begin{equation}
    \begin{split}
        G_{xx}& \simeq 4k_y^2 v_x^2 v_y^2/(k_x^2 v_x^2 + k_y^2 v_y^2)^{3/2}, \\
        G_{yy}& \simeq 4k_x^2 v_x^2 v_y^2/(k_x^2 v_x^2 + k_y^2 v_y^2)^{3/2}, \\
        G_{xy}& \simeq -4 k_x k_y v_x^2 v_y^2/(k_x^2 v_x^2 + k_y^2 v_y^2)^{3/2}. 
    \end{split}
\end{equation}
where it is clear that $G_{xx} \to G_{xx}$ for $k_y \to -k_y$, $G_{yy} \to G_{yy}$ for $k_x \to -k_x$, and $G_{xy} \to -G_{xy}$ for $(k_x,k_y) \to (-k_x,k_y)$ or $(k_x,k_y) \to (k_x,-k_y)$.

Now to explore the warping effect on the BCP tensor we need to look away from the $\Gamma$-point because the warping term  $\propto k^3$ acquires very small value close to $\Gamma$-point (note that, this effect vanishes at $\Gamma$-point). We find that the magnitude of the BCP components increases with increasing $\lambda$ away from the $\Gamma$-point. In particular, the $G_{xx}$ captures snowflake like structure whereas for $G_{yy}$ and $G_{xy}$ show the sharp tips and inward curves caused by the warping effect.
To be precise, $G_{yy}$ and $G_{xy}$ show  quadrupolar features as caused by the warping effect (see middle and lower panel in Fig.~~\ref{BCP_Tensor}). The four legs of the quadrupole acquire same sign for $G_{yy}$ while sign changes between two consecutive legs for $G_{xy}$. The quadrupolar structure can be caused by $k_x k_y$-product terms in $G_{yy}$ and $G_{xy}$.
This nature is substantially different  compared to BC which  always shows snowflake-like structure with any finite warping strength. It is important to note that the tilt parameter does not affect the BCP tensor and it only causes the anisotropic shifting of the Fermi surface along the tilt direction. This  is also depicted in lower panel Fig.~\ref{BCP_Tensor}.
Notice that the behavior of BCP tensor close to 
$\Gamma$-point remains insensitive to the warping strength as shown in the insets Fig.~\ref{BCP_Tensor}. This obseration can be analytically understood from the expression of BCP tensor derived above.  
We want to mention that 
the variation in the BCP magnitude and  the anisotropy in the Fermi surface can  either simultaneously or separately
impact the net TOH responses.

With the BCP tensor in hand, we will now calculate the TOH conductivity using Eq.~(\ref{eqcondtens}). We would like to point out that since the system is invariant under mirror symmetry $M_x$, the tensor components involving odd number of $x$ and $y$ (e.g., $\chi_{xxxy}$, $\chi_{yxxx}$) vanish. To explore the angular dependence of the TOHE, we consider the applied electric field $\mathbf{E}=E(\cos \theta, \sin \theta,0)$,  making an angle $\theta$ with the $x$-axis (that is $\perp$ to the mirror line in the current study). Since we are interested in TOH response, the transverse third-order conductivity can be written as 
\begin{eqnarray}
\chi_{3}^{\rm H}(\theta)=J_{3}^{\rm H}/E^3
\end{eqnarray}
where $j_{3}^{\rm H}=\mathbf{j}_{3} \cdot (\hat{z} \times \mathbf{E}) $ is the current flowing perpendicular to the applied electric field. Now using the above equation, the explicit expression of $\chi_{3}^{\rm H}(\theta)$ can be obtained as
\begin{eqnarray}
&&\chi_{3}^{\rm H}(\theta) \nonumber \\
&&=(3\chi_{21}-\chi_{11})\sin \theta \cos^3 \theta - (3\chi_{12}-\chi_{22})\sin^3 \theta \cos \theta \nonumber \\
\end{eqnarray}
where $\chi_{11}=\chi_{xxxx}$, $\chi_{22}=\chi_{yyyy}$, $\chi_{12}=\frac{1}{3}(\chi_{xxyy}+\chi_{xyxy}+\chi_{xyyx})$ and $\chi_{21}=\frac{1}{3}(\chi_{yyxx}+\chi_{yxyx}+\chi_{yxxy})$. It is now clear from the expression that the TOH current vanishes along (with $\theta=\pi/2$) and perpendicular (with $\theta=0$) to the mirror line. 

In the absence of tilting, the $C_{3}$ symmetry is preserved and the Fermi surface is circular when $v_x=v_y$. Therefore, the $\chi_{3}^{\rm H}(\theta)$ vanishes due to symmetry constraint. Now to get the finite contribution of TOHE, one needs to break the $C_{3}$ symmetry. This can be achieved by introducing tilt parameter $\omega$ where Dirac cone remains gapless. Alternatively, one can  consider anisotropic Dirac velocities i.e., $v_x \neq v_y$, preserving TRS, to break  $C_{3}$ symmetry. Note that hexagonal warping term does not break $C_{3}$ symmetry. The variation of $\chi_{3}^{\rm H}$ as a function of $\theta$ for different tilt strength is shown in the top panel  Fig.~\ref{TOHE_Tensor}. It is important to note that the BCP is singular at $k=0$ for the gapless surface states of TIs which is very much similar to that of the Berry curvature and Berry curvature dipole when two bands are approaching each other (around the degenerate points). However, one can consider an infinitesimal constant in the gap in the numerical calculations to avoid the singularity.
We find that although the qualitative changes remain same, the magnitude of the conductivity enhances with increasing tilt strength. This happens because the stronger band tilt provides a more anisotropically warped Fermi surface, which straightforwardly modifies the net contribution of the current (see Eqs.~(\ref{dis_fun})-(\ref{eqcondtens})). Next we show the variation of $\chi_{3}^{\rm H}(\theta)$ 
in middle panel  Fig.~\ref{TOHE_Tensor}
for different warping strength $\lambda$ with chemical potential $\mu=0.25$ eV and $\omega=0.2$. Clearly, the magnitude of the TOH conductivity increases with increasing the warping strength which can be visualized from the evaluation of the BCP tensor with warping. Finally, we investigate the effect of doping on $\chi_{3}^{\rm H}(\theta)$. We find that the peak positions of $\chi_{3}^{\rm H}(\theta)$ shifts toward lower angle whereas the dip positions shift toward higher angle. Moreover, the response becomes substantially strong  for $\mu > 0.2~eV$
as the warping effect of the surface Dirac cone is only apparent above some chemical potential threshold~\cite{Fu09}.

It is interesting to note that for 
$C_3$ broken TRS preserved case with 
anisotropic non-tilted 2D Dirac cone (i.e., $v_x \neq v_y$), the BCD induced as well as disorder mediated second-order Hall conductivity vanishes. This is due to the fact that such responses are  proportional to the tilt factor $\omega$ associated with identity term \cite{sodemann15,Nandy2019}. By constrast,   BCP induced TOHE survives. Noticeably, the TOHE is not directly related to the tilt factor as the BCP tensor is insensitive with respect to the above. Therefore, TOH response could be the leading-order response for TI systems consisting non-tilted anisotropic Dirac cone on the surface states because of the presence of TR symmetry.


\section{Discussion and conclusions}
\label{conc}

In summary, within the framework of quasiclassical Boltzmann theory, we consider the general expression of TOH current. We show that in a TRS invariant system, the TOH current, appearing due to the BCP tensor, is   proportional to relaxation time $\tau$. Using the symmetry arguments, we  establish that such BCP induced TOH response can become the leading-order response in the surface states of 3D TI in the absence of tilting of the Dirac cone. This provides  a direct  check for TOHE  in experiments. 

We explore the effect of warping strength, chemical doping and tilt on TOH response in this system. We find that the magnitude of the TOH conductivity can be  enhanced significantly by increasing the hexagonal warping strength $\lambda$ which is inherently present in the surface states of 3D TI such as Bi$_2$Te$_3$ (see Fig.~\ref{TOHE_Tensor}). This is related to the fact that the magnitudes of the BCP tensor-components, shown in Fig.~\ref{BCP_Tensor},  increase with increasing $\lambda$ away from the $\Gamma$-point.  Our study also reveals that  the tilt strength has no effect on BCP tensor unlike the warping parameter, however, tilt parameter can also tune TOHE by enhancing the anisotropy of the Fermi surface. As the BCP is closely dependent on the warping strength, the effective TOHE is also shown to exibit more significant signals at the higher chemical potential where the warping becomes evident for the surface states. 

We here comment on the possible  experimental  predictions as fas as the values of TOH are concerned. For the 20 QL-thick ($t \sim 20~nm$) TI material (e.g. $\mathrm{Bi_2Te_3}$) fabricated in the conventional Hall bar geometry with size $l \times w =100~\mu m \times 20~\mu m$~\cite{Steven19,He_2018}, 
considering the resistivity $\rho = 700~\mu \Omega \cdot cm$, scattering time $\tau =5.86 \times 10^{-13}~s$, a current drive with magnitude of $I_{0} =0.6~mA$ can provide an electric field $E =105 V\cdot cm^{-1}$. When the field angle $\theta$ is around $\theta\sim \pi/6$ (see the top panel of Fig.~2), the induced TOH voltage $U_H \propto \chi^H_3 E^3 $ at $\mu =0.25~ eV$ can be estimated to be $U_H \sim 14.02~ \mu V$ when a weak band tilt $w/v_x =0.05$ is present. A moderate band tilt $w/v_x =0.20 $ can even lead to a TOH voltage in the order of $U_H \sim 0.129~ mV$. Similar estimations can be also made for TI surface Dirac states when it is non-tilted but with slightly anisotropic Fermi velocities ($v_x \neq v_y$).

Moreover, we also discuss in detail that how to separate the third-order Hall response from linear and second-order effect in experiments. Since the dictionary of topological materials possesing warping and tilt at the same time is diverse, our work opens an avenue for searching TOHE in various topological systems.
It is important to note that the effective mass of the material can reduce the warping effect, fortunately, as have been demonstrated in experiments, the warping effect is indeed present for the surface states in TIs. Specifically, the suppression coming from the effective mass on the warping effect (as well as the band tilt) is found to be much less in $\mathrm{Bi_{2}Te_{3}}$ than in other TI~\cite{tss_2010prl,Ando_2014prb,kimura_2018prb,NLNE_czeng_2022}. Therefore, the BCP driven  nonlinear transport, discussed in our work, is expected to be experimentally viable for $\mathrm{Bi_{2}Te_{3}}$.

 \section{Acknowledgements}
The work at Los Alamos National Laboratory was carried out under the auspices
of the US Department of Energy (DOE) National
Nuclear Security Administration under Contract No.
89233218CNA000001. It was supported by the LANL
LDRD Program, and in part by the Center for Integrated Nanotechnologies, a DOE BES user facility, in partnership with the LANL Institutional Computing Program for computational resources. S.K.D was partially supported by the Startup grant of Bitan Roy from Lehigh University. C. Z. acknowledges support from NSFC (Grant No. 12104043), the fellowship of the China Postdoctoral Science Foundation (Grant No. 2021M690409) and the National Key R$\&$D Program of China (Grant No. 2020YFA0308800).

\bibliography{TOHE_2D}

\begin{thebibliography}{66}%
\makeatletter
\providecommand \@ifxundefined [1]{%
 \@ifx{#1\undefined}
}%
\providecommand \@ifnum [1]{%
 \ifnum #1\expandafter \@firstoftwo
 \else \expandafter \@secondoftwo
 \fi
}%
\providecommand \@ifx [1]{%
 \ifx #1\expandafter \@firstoftwo
 \else \expandafter \@secondoftwo
 \fi
}%
\providecommand \natexlab [1]{#1}%
\providecommand \enquote  [1]{``#1''}%
\providecommand \bibnamefont  [1]{#1}%
\providecommand \bibfnamefont [1]{#1}%
\providecommand \citenamefont [1]{#1}%
\providecommand \href@noop [0]{\@secondoftwo}%
\providecommand \href [0]{\begingroup \@sanitize@url \@href}%
\providecommand \@href[1]{\@@startlink{#1}\@@href}%
\providecommand \@@href[1]{\endgroup#1\@@endlink}%
\providecommand \@sanitize@url [0]{\catcode `\\12\catcode `\$12\catcode
  `\&12\catcode `\#12\catcode `\^12\catcode `\_12\catcode `\%12\relax}%
\providecommand \@@startlink[1]{}%
\providecommand \@@endlink[0]{}%
\providecommand \url  [0]{\begingroup\@sanitize@url \@url }%
\providecommand \@url [1]{\endgroup\@href {#1}{\urlprefix }}%
\providecommand \urlprefix  [0]{URL }%
\providecommand \Eprint [0]{\href }%
\providecommand \doibase [0]{http://dx.doi.org/}%
\providecommand \selectlanguage [0]{\@gobble}%
\providecommand \bibinfo  [0]{\@secondoftwo}%
\providecommand \bibfield  [0]{\@secondoftwo}%
\providecommand \translation [1]{[#1]}%
\providecommand \BibitemOpen [0]{}%
\providecommand \bibitemStop [0]{}%
\providecommand \bibitemNoStop [0]{.\EOS\space}%
\providecommand \EOS [0]{\spacefactor3000\relax}%
\providecommand \BibitemShut  [1]{\csname bibitem#1\endcsname}%
\let\auto@bib@innerbib\@empty
\bibitem [{\citenamefont {Hasan}\ and\ \citenamefont {Kane}(2010)}]{Hasan2010}%
  \BibitemOpen
  \bibfield  {author} {\bibinfo {author} {\bibfnamefont {M.~Z.}\ \bibnamefont
  {Hasan}}\ and\ \bibinfo {author} {\bibfnamefont {C.~L.}\ \bibnamefont
  {Kane}},\ }\href {\doibase 10.1103/RevModPhys.82.3045} {\bibfield  {journal}
  {\bibinfo  {journal} {Rev. Mod. Phys.}\ }\textbf {\bibinfo {volume} {82}},\
  \bibinfo {pages} {3045} (\bibinfo {year} {2010})}\BibitemShut {NoStop}%
\bibitem [{\citenamefont {Hasan}\ and\ \citenamefont
  {Moore}(2011)}]{Hasan2011}%
  \BibitemOpen
  \bibfield  {author} {\bibinfo {author} {\bibfnamefont {M.~Z.}\ \bibnamefont
  {Hasan}}\ and\ \bibinfo {author} {\bibfnamefont {J.~E.}\ \bibnamefont
  {Moore}},\ }\href {\doibase 10.1146/annurev-conmatphys-062910-140432}
  {\bibfield  {journal} {\bibinfo  {journal} {Annual Review of Condensed Matter
  Physics}\ }\textbf {\bibinfo {volume} {2}},\ \bibinfo {pages} {55} (\bibinfo
  {year} {2011})}\BibitemShut {NoStop}%
\bibitem [{\citenamefont {Qi}\ and\ \citenamefont {Zhang}(2011)}]{Qi2011}%
  \BibitemOpen
  \bibfield  {author} {\bibinfo {author} {\bibfnamefont {X.-L.}\ \bibnamefont
  {Qi}}\ and\ \bibinfo {author} {\bibfnamefont {S.-C.}\ \bibnamefont {Zhang}},\
  }\href {\doibase 10.1103/RevModPhys.83.1057} {\bibfield  {journal} {\bibinfo
  {journal} {Rev. Mod. Phys.}\ }\textbf {\bibinfo {volume} {83}},\ \bibinfo
  {pages} {1057} (\bibinfo {year} {2011})}\BibitemShut {NoStop}%
\bibitem [{\citenamefont {Pesin}\ and\ \citenamefont
  {MacDonald}(2012)}]{Pesin2012}%
  \BibitemOpen
  \bibfield  {author} {\bibinfo {author} {\bibfnamefont {D.}~\bibnamefont
  {Pesin}}\ and\ \bibinfo {author} {\bibfnamefont {A.~H.}\ \bibnamefont
  {MacDonald}},\ }\href {\doibase 10.1038/nmat3305} {\bibfield  {journal}
  {\bibinfo  {journal} {Nature Materials}\ }\textbf {\bibinfo {volume} {11}},\
  \bibinfo {pages} {409} (\bibinfo {year} {2012})}\BibitemShut {NoStop}%
\bibitem [{\citenamefont {Ando}(2013)}]{Ando2013}%
  \BibitemOpen
  \bibfield  {author} {\bibinfo {author} {\bibfnamefont {Y.}~\bibnamefont
  {Ando}},\ }\href {\doibase 10.7566/jpsj.82.102001} {\bibfield  {journal}
  {\bibinfo  {journal} {Journal of the Physical Society of Japan}\ }\textbf
  {\bibinfo {volume} {82}},\ \bibinfo {pages} {102001} (\bibinfo {year}
  {2013})}\BibitemShut {NoStop}%
\bibitem [{\citenamefont {Klitzing}\ \emph {et~al.}(1980)\citenamefont
  {Klitzing}, \citenamefont {Dorda},\ and\ \citenamefont
  {Pepper}}]{klitzing80}%
  \BibitemOpen
  \bibfield  {author} {\bibinfo {author} {\bibfnamefont {K.~v.}\ \bibnamefont
  {Klitzing}}, \bibinfo {author} {\bibfnamefont {G.}~\bibnamefont {Dorda}}, \
  and\ \bibinfo {author} {\bibfnamefont {M.}~\bibnamefont {Pepper}},\ }\href
  {\doibase 10.1103/PhysRevLett.45.494} {\bibfield  {journal} {\bibinfo
  {journal} {Phys. Rev. Lett.}\ }\textbf {\bibinfo {volume} {45}},\ \bibinfo
  {pages} {494} (\bibinfo {year} {1980})}\BibitemShut {NoStop}%
\bibitem [{\citenamefont {Kane}\ and\ \citenamefont {Mele}(2005)}]{kane05}%
  \BibitemOpen
  \bibfield  {author} {\bibinfo {author} {\bibfnamefont {C.~L.}\ \bibnamefont
  {Kane}}\ and\ \bibinfo {author} {\bibfnamefont {E.~J.}\ \bibnamefont
  {Mele}},\ }\href {\doibase 10.1103/PhysRevLett.95.226801} {\bibfield
  {journal} {\bibinfo  {journal} {Phys. Rev. Lett.}\ }\textbf {\bibinfo
  {volume} {95}},\ \bibinfo {pages} {226801} (\bibinfo {year}
  {2005})}\BibitemShut {NoStop}%
\bibitem [{\citenamefont {Haldane}(1988)}]{haldane88}%
  \BibitemOpen
  \bibfield  {author} {\bibinfo {author} {\bibfnamefont {F.~D.~M.}\
  \bibnamefont {Haldane}},\ }\href {\doibase 10.1103/PhysRevLett.61.2015}
  {\bibfield  {journal} {\bibinfo  {journal} {Phys. Rev. Lett.}\ }\textbf
  {\bibinfo {volume} {61}},\ \bibinfo {pages} {2015} (\bibinfo {year}
  {1988})}\BibitemShut {NoStop}%
\bibitem [{\citenamefont {Liu}\ \emph {et~al.}(2016)\citenamefont {Liu},
  \citenamefont {Zhang},\ and\ \citenamefont {Qi}}]{liu16}%
  \BibitemOpen
  \bibfield  {author} {\bibinfo {author} {\bibfnamefont {C.-X.}\ \bibnamefont
  {Liu}}, \bibinfo {author} {\bibfnamefont {S.-C.}\ \bibnamefont {Zhang}}, \
  and\ \bibinfo {author} {\bibfnamefont {X.-L.}\ \bibnamefont {Qi}},\ }\href
  {\doibase 10.1146/annurev-conmatphys-031115-011417} {\bibfield  {journal}
  {\bibinfo  {journal} {Annual Review of Condensed Matter Physics}\ }\textbf
  {\bibinfo {volume} {7}},\ \bibinfo {pages} {301} (\bibinfo {year}
  {2016})}\BibitemShut {NoStop}%
\bibitem [{\citenamefont {Xiao}\ \emph {et~al.}(2006)\citenamefont {Xiao},
  \citenamefont {Yao}, \citenamefont {Fang},\ and\ \citenamefont
  {Niu}}]{Xiaonernst}%
  \BibitemOpen
  \bibfield  {author} {\bibinfo {author} {\bibfnamefont {D.}~\bibnamefont
  {Xiao}}, \bibinfo {author} {\bibfnamefont {Y.}~\bibnamefont {Yao}}, \bibinfo
  {author} {\bibfnamefont {Z.}~\bibnamefont {Fang}}, \ and\ \bibinfo {author}
  {\bibfnamefont {Q.}~\bibnamefont {Niu}},\ }\href {\doibase
  10.1103/PhysRevLett.97.026603} {\bibfield  {journal} {\bibinfo  {journal}
  {Phys. Rev. Lett.}\ }\textbf {\bibinfo {volume} {97}},\ \bibinfo {pages}
  {026603} (\bibinfo {year} {2006})}\BibitemShut {NoStop}%
\bibitem [{Ber(1984)}]{Berry1984}%
  \BibitemOpen
  \href {\doibase 10.1098/rspa.1984.0023} {\bibfield  {journal} {\bibinfo
  {journal} {Proceedings of the Royal Society of London. A. Mathematical and
  Physical Sciences}\ }\textbf {\bibinfo {volume} {392}},\ \bibinfo {pages}
  {45} (\bibinfo {year} {1984})}\BibitemShut {NoStop}%
\bibitem [{\citenamefont {Xiao}\ \emph {et~al.}(2010)\citenamefont {Xiao},
  \citenamefont {Chang},\ and\ \citenamefont {Niu}}]{Niu_2010}%
  \BibitemOpen
  \bibfield  {author} {\bibinfo {author} {\bibfnamefont {D.}~\bibnamefont
  {Xiao}}, \bibinfo {author} {\bibfnamefont {M.-C.}\ \bibnamefont {Chang}}, \
  and\ \bibinfo {author} {\bibfnamefont {Q.}~\bibnamefont {Niu}},\ }\href
  {\doibase 10.1103/RevModPhys.82.1959} {\bibfield  {journal} {\bibinfo
  {journal} {Rev. Mod. Phys.}\ }\textbf {\bibinfo {volume} {82}},\ \bibinfo
  {pages} {1959} (\bibinfo {year} {2010})}\BibitemShut {NoStop}%
\bibitem [{\citenamefont {Sodemann}\ and\ \citenamefont
  {Fu}(2015)}]{sodemann15}%
  \BibitemOpen
  \bibfield  {author} {\bibinfo {author} {\bibfnamefont {I.}~\bibnamefont
  {Sodemann}}\ and\ \bibinfo {author} {\bibfnamefont {L.}~\bibnamefont {Fu}},\
  }\href {\doibase 10.1103/PhysRevLett.115.216806} {\bibfield  {journal}
  {\bibinfo  {journal} {Phys. Rev. Lett.}\ }\textbf {\bibinfo {volume} {115}},\
  \bibinfo {pages} {216806} (\bibinfo {year} {2015})}\BibitemShut {NoStop}%
\bibitem [{\citenamefont {Samal}\ \emph {et~al.}(2021)\citenamefont {Samal},
  \citenamefont {Nandy},\ and\ \citenamefont {Saha}}]{Nandy_21}%
  \BibitemOpen
  \bibfield  {author} {\bibinfo {author} {\bibfnamefont {S.~S.}\ \bibnamefont
  {Samal}}, \bibinfo {author} {\bibfnamefont {S.}~\bibnamefont {Nandy}}, \ and\
  \bibinfo {author} {\bibfnamefont {K.}~\bibnamefont {Saha}},\ }\href {\doibase
  10.1103/PhysRevB.103.L201202} {\bibfield  {journal} {\bibinfo  {journal}
  {Phys. Rev. B}\ }\textbf {\bibinfo {volume} {103}},\ \bibinfo {pages}
  {L201202} (\bibinfo {year} {2021})}\BibitemShut {NoStop}%
\bibitem [{\citenamefont {Nandy}\ and\ \citenamefont
  {Sodemann}(2019)}]{Nandy2019}%
  \BibitemOpen
  \bibfield  {author} {\bibinfo {author} {\bibfnamefont {S.}~\bibnamefont
  {Nandy}}\ and\ \bibinfo {author} {\bibfnamefont {I.}~\bibnamefont
  {Sodemann}},\ }\href {\doibase 10.1103/PhysRevB.100.195117} {\bibfield
  {journal} {\bibinfo  {journal} {Phys. Rev. B}\ }\textbf {\bibinfo {volume}
  {100}},\ \bibinfo {pages} {195117} (\bibinfo {year} {2019})}\BibitemShut
  {NoStop}%
\bibitem [{\citenamefont {Matsyshyn}\ and\ \citenamefont
  {Sodemann}(2019)}]{Matsyshyn2019}%
  \BibitemOpen
  \bibfield  {author} {\bibinfo {author} {\bibfnamefont {O.}~\bibnamefont
  {Matsyshyn}}\ and\ \bibinfo {author} {\bibfnamefont {I.}~\bibnamefont
  {Sodemann}},\ }\href {\doibase 10.1103/PhysRevLett.123.246602} {\bibfield
  {journal} {\bibinfo  {journal} {Phys. Rev. Lett.}\ }\textbf {\bibinfo
  {volume} {123}},\ \bibinfo {pages} {246602} (\bibinfo {year}
  {2019})}\BibitemShut {NoStop}%
\bibitem [{\citenamefont {Das}\ \emph {et~al.}(2021)\citenamefont {Das},
  \citenamefont {Nag},\ and\ \citenamefont {Nandy}}]{Das21}%
  \BibitemOpen
  \bibfield  {author} {\bibinfo {author} {\bibfnamefont {S.~K.}\ \bibnamefont
  {Das}}, \bibinfo {author} {\bibfnamefont {T.}~\bibnamefont {Nag}}, \ and\
  \bibinfo {author} {\bibfnamefont {S.}~\bibnamefont {Nandy}},\ }\href
  {\doibase 10.1103/PhysRevB.104.115420} {\bibfield  {journal} {\bibinfo
  {journal} {Phys. Rev. B}\ }\textbf {\bibinfo {volume} {104}},\ \bibinfo
  {pages} {115420} (\bibinfo {year} {2021})}\BibitemShut {NoStop}%
\bibitem [{\citenamefont {Nag}\ and\ \citenamefont {Kennes}(2022)}]{Nag22a}%
  \BibitemOpen
  \bibfield  {author} {\bibinfo {author} {\bibfnamefont {T.}~\bibnamefont
  {Nag}}\ and\ \bibinfo {author} {\bibfnamefont {D.~M.}\ \bibnamefont
  {Kennes}},\ }\href {\doibase 10.1103/PhysRevB.105.214307} {\bibfield
  {journal} {\bibinfo  {journal} {Phys. Rev. B}\ }\textbf {\bibinfo {volume}
  {105}},\ \bibinfo {pages} {214307} (\bibinfo {year} {2022})}\BibitemShut
  {NoStop}%
\bibitem [{\citenamefont {Sadhukhan}\ and\ \citenamefont
  {Nag}(2021{\natexlab{a}})}]{Sadhukhan21a}%
  \BibitemOpen
  \bibfield  {author} {\bibinfo {author} {\bibfnamefont {B.}~\bibnamefont
  {Sadhukhan}}\ and\ \bibinfo {author} {\bibfnamefont {T.}~\bibnamefont
  {Nag}},\ }\href {\doibase 10.1103/PhysRevB.104.245122} {\bibfield  {journal}
  {\bibinfo  {journal} {Phys. Rev. B}\ }\textbf {\bibinfo {volume} {104}},\
  \bibinfo {pages} {245122} (\bibinfo {year} {2021}{\natexlab{a}})}\BibitemShut
  {NoStop}%
\bibitem [{\citenamefont {Sadhukhan}\ and\ \citenamefont
  {Nag}(2021{\natexlab{b}})}]{Sadhukhan21b}%
  \BibitemOpen
  \bibfield  {author} {\bibinfo {author} {\bibfnamefont {B.}~\bibnamefont
  {Sadhukhan}}\ and\ \bibinfo {author} {\bibfnamefont {T.}~\bibnamefont
  {Nag}},\ }\href {\doibase 10.1103/PhysRevB.103.144308} {\bibfield  {journal}
  {\bibinfo  {journal} {Phys. Rev. B}\ }\textbf {\bibinfo {volume} {103}},\
  \bibinfo {pages} {144308} (\bibinfo {year} {2021}{\natexlab{b}})}\BibitemShut
  {NoStop}%
\bibitem [{\citenamefont {Liu}\ \emph {et~al.}(2021)\citenamefont {Liu},
  \citenamefont {Zhao}, \citenamefont {Huang}, \citenamefont {Feng},
  \citenamefont {Xiao}, \citenamefont {Wu}, \citenamefont {Lai}, \citenamefont
  {bo~Gao},\ and\ \citenamefont {Yang}}]{Liu2021}%
  \BibitemOpen
  \bibfield  {author} {\bibinfo {author} {\bibfnamefont {H.}~\bibnamefont
  {Liu}}, \bibinfo {author} {\bibfnamefont {J.}~\bibnamefont {Zhao}}, \bibinfo
  {author} {\bibfnamefont {Y.}~\bibnamefont {Huang}}, \bibinfo {author}
  {\bibfnamefont {X.}~\bibnamefont {Feng}}, \bibinfo {author} {\bibfnamefont
  {C.}~\bibnamefont {Xiao}}, \bibinfo {author} {\bibfnamefont {W.}~\bibnamefont
  {Wu}}, \bibinfo {author} {\bibfnamefont {S.}~\bibnamefont {Lai}}, \bibinfo
  {author} {\bibfnamefont {W.}~\bibnamefont {bo~Gao}}, \ and\ \bibinfo {author}
  {\bibfnamefont {S.~A.}\ \bibnamefont {Yang}},\ }\href@noop {} {} (\bibinfo
  {year} {2021}),\ \Eprint {http://arxiv.org/abs/arXiv:2106.04931}
  {arXiv:2106.04931} \BibitemShut {NoStop}%
\bibitem [{\citenamefont {Wei}\ \emph {et~al.}(2022)\citenamefont {Wei},
  \citenamefont {Xiang}, \citenamefont {Wang}, \citenamefont {Xu},\ and\
  \citenamefont {Wang}}]{Wei_2022}%
  \BibitemOpen
  \bibfield  {author} {\bibinfo {author} {\bibfnamefont {M.}~\bibnamefont
  {Wei}}, \bibinfo {author} {\bibfnamefont {L.}~\bibnamefont {Xiang}}, \bibinfo
  {author} {\bibfnamefont {L.}~\bibnamefont {Wang}}, \bibinfo {author}
  {\bibfnamefont {F.}~\bibnamefont {Xu}}, \ and\ \bibinfo {author}
  {\bibfnamefont {J.}~\bibnamefont {Wang}},\ }\href {\doibase
  10.1103/PhysRevB.106.035307} {\bibfield  {journal} {\bibinfo  {journal}
  {Phys. Rev. B}\ }\textbf {\bibinfo {volume} {106}},\ \bibinfo {pages}
  {035307} (\bibinfo {year} {2022})}\BibitemShut {NoStop}%
\bibitem [{\citenamefont {Zhang}\ \emph {et~al.}(2022)\citenamefont {Zhang},
  \citenamefont {Zhu},\ and\ \citenamefont {Su}}]{Su_2022}%
  \BibitemOpen
  \bibfield  {author} {\bibinfo {author} {\bibfnamefont {Z.-f.}\ \bibnamefont
  {Zhang}}, \bibinfo {author} {\bibfnamefont {Z.-G.}\ \bibnamefont {Zhu}}, \
  and\ \bibinfo {author} {\bibfnamefont {G.}~\bibnamefont {Su}},\ }\href
  {\doibase 10.48550/ARXIV.2209.08556} {\enquote {\bibinfo {title} {Symmetry
  dictionary on charge and spin nonlinear responses for all magnetic point
  groups with nontrivial topological nature},}\ } (\bibinfo {year}
  {2022})\BibitemShut {NoStop}%
\bibitem [{\citenamefont {Gao}\ \emph {et~al.}(2015)\citenamefont {Gao},
  \citenamefont {Yang},\ and\ \citenamefont {Niu}}]{Gao2015}%
  \BibitemOpen
  \bibfield  {author} {\bibinfo {author} {\bibfnamefont {Y.}~\bibnamefont
  {Gao}}, \bibinfo {author} {\bibfnamefont {S.~A.}\ \bibnamefont {Yang}}, \
  and\ \bibinfo {author} {\bibfnamefont {Q.}~\bibnamefont {Niu}},\ }\href
  {\doibase 10.1103/PhysRevB.91.214405} {\bibfield  {journal} {\bibinfo
  {journal} {Phys. Rev. B}\ }\textbf {\bibinfo {volume} {91}},\ \bibinfo
  {pages} {214405} (\bibinfo {year} {2015})}\BibitemShut {NoStop}%
\bibitem [{\citenamefont {Low}\ \emph {et~al.}(2015)\citenamefont {Low},
  \citenamefont {Jiang},\ and\ \citenamefont {Guinea}}]{Low2015}%
  \BibitemOpen
  \bibfield  {author} {\bibinfo {author} {\bibfnamefont {T.}~\bibnamefont
  {Low}}, \bibinfo {author} {\bibfnamefont {Y.}~\bibnamefont {Jiang}}, \ and\
  \bibinfo {author} {\bibfnamefont {F.}~\bibnamefont {Guinea}},\ }\href
  {\doibase 10.1103/PhysRevB.92.235447} {\bibfield  {journal} {\bibinfo
  {journal} {Phys. Rev. B}\ }\textbf {\bibinfo {volume} {92}},\ \bibinfo
  {pages} {235447} (\bibinfo {year} {2015})}\BibitemShut {NoStop}%
\bibitem [{\citenamefont {Zhang}\ \emph
  {et~al.}(2018{\natexlab{a}})\citenamefont {Zhang}, \citenamefont {Sun},\ and\
  \citenamefont {Yan}}]{Zhang2018}%
  \BibitemOpen
  \bibfield  {author} {\bibinfo {author} {\bibfnamefont {Y.}~\bibnamefont
  {Zhang}}, \bibinfo {author} {\bibfnamefont {Y.}~\bibnamefont {Sun}}, \ and\
  \bibinfo {author} {\bibfnamefont {B.}~\bibnamefont {Yan}},\ }\href {\doibase
  10.1103/PhysRevB.97.041101} {\bibfield  {journal} {\bibinfo  {journal} {Phys.
  Rev. B}\ }\textbf {\bibinfo {volume} {97}},\ \bibinfo {pages} {041101}
  (\bibinfo {year} {2018}{\natexlab{a}})}\BibitemShut {NoStop}%
\bibitem [{\citenamefont {Zeng}\ \emph {et~al.}(2021)\citenamefont {Zeng},
  \citenamefont {Nandy},\ and\ \citenamefont {Tewari}}]{BCD_2021}%
  \BibitemOpen
  \bibfield  {author} {\bibinfo {author} {\bibfnamefont {C.}~\bibnamefont
  {Zeng}}, \bibinfo {author} {\bibfnamefont {S.}~\bibnamefont {Nandy}}, \ and\
  \bibinfo {author} {\bibfnamefont {S.}~\bibnamefont {Tewari}},\ }\href
  {\doibase 10.1103/PhysRevB.103.245119} {\bibfield  {journal} {\bibinfo
  {journal} {Phys. Rev. B}\ }\textbf {\bibinfo {volume} {103}},\ \bibinfo
  {pages} {245119} (\bibinfo {year} {2021})}\BibitemShut {NoStop}%
\bibitem [{\citenamefont {Facio}\ \emph {et~al.}(2018)\citenamefont {Facio},
  \citenamefont {Efremov}, \citenamefont {Koepernik}, \citenamefont {You},
  \citenamefont {Sodemann},\ and\ \citenamefont {van~den Brink}}]{Facio2018}%
  \BibitemOpen
  \bibfield  {author} {\bibinfo {author} {\bibfnamefont {J.~I.}\ \bibnamefont
  {Facio}}, \bibinfo {author} {\bibfnamefont {D.}~\bibnamefont {Efremov}},
  \bibinfo {author} {\bibfnamefont {K.}~\bibnamefont {Koepernik}}, \bibinfo
  {author} {\bibfnamefont {J.-S.}\ \bibnamefont {You}}, \bibinfo {author}
  {\bibfnamefont {I.}~\bibnamefont {Sodemann}}, \ and\ \bibinfo {author}
  {\bibfnamefont {J.}~\bibnamefont {van~den Brink}},\ }\href {\doibase
  10.1103/PhysRevLett.121.246403} {\bibfield  {journal} {\bibinfo  {journal}
  {Phys. Rev. Lett.}\ }\textbf {\bibinfo {volume} {121}},\ \bibinfo {pages}
  {246403} (\bibinfo {year} {2018})}\BibitemShut {NoStop}%
\bibitem [{\citenamefont {Du}\ \emph {et~al.}(2018)\citenamefont {Du},
  \citenamefont {Wang}, \citenamefont {Lu},\ and\ \citenamefont
  {Xie}}]{Du2018}%
  \BibitemOpen
  \bibfield  {author} {\bibinfo {author} {\bibfnamefont {Z.~Z.}\ \bibnamefont
  {Du}}, \bibinfo {author} {\bibfnamefont {C.~M.}\ \bibnamefont {Wang}},
  \bibinfo {author} {\bibfnamefont {H.-Z.}\ \bibnamefont {Lu}}, \ and\ \bibinfo
  {author} {\bibfnamefont {X.~C.}\ \bibnamefont {Xie}},\ }\href {\doibase
  10.1103/PhysRevLett.121.266601} {\bibfield  {journal} {\bibinfo  {journal}
  {Phys. Rev. Lett.}\ }\textbf {\bibinfo {volume} {121}},\ \bibinfo {pages}
  {266601} (\bibinfo {year} {2018})}\BibitemShut {NoStop}%
\bibitem [{\citenamefont {You}\ \emph {et~al.}(2018)\citenamefont {You},
  \citenamefont {Fang}, \citenamefont {Xu}, \citenamefont {Kaxiras},\ and\
  \citenamefont {Low}}]{You2018}%
  \BibitemOpen
  \bibfield  {author} {\bibinfo {author} {\bibfnamefont {J.-S.}\ \bibnamefont
  {You}}, \bibinfo {author} {\bibfnamefont {S.}~\bibnamefont {Fang}}, \bibinfo
  {author} {\bibfnamefont {S.-Y.}\ \bibnamefont {Xu}}, \bibinfo {author}
  {\bibfnamefont {E.}~\bibnamefont {Kaxiras}}, \ and\ \bibinfo {author}
  {\bibfnamefont {T.}~\bibnamefont {Low}},\ }\href {\doibase
  10.1103/PhysRevB.98.121109} {\bibfield  {journal} {\bibinfo  {journal} {Phys.
  Rev. B}\ }\textbf {\bibinfo {volume} {98}},\ \bibinfo {pages} {121109}
  (\bibinfo {year} {2018})}\BibitemShut {NoStop}%
\bibitem [{\citenamefont {Ma}\ \emph {et~al.}(2018)\citenamefont {Ma},
  \citenamefont {Xu}, \citenamefont {Shen}, \citenamefont {MacNeill},
  \citenamefont {Fatemi}, \citenamefont {Chang}, \citenamefont {Valdivia},
  \citenamefont {Wu}, \citenamefont {Du}, \citenamefont {Hsu}, \citenamefont
  {Fang}, \citenamefont {Gibson}, \citenamefont {Watanabe}, \citenamefont
  {Taniguchi}, \citenamefont {Cava}, \citenamefont {Kaxiras}, \citenamefont
  {Lu}, \citenamefont {Lin}, \citenamefont {Fu}, \citenamefont {Gedik},\ and\
  \citenamefont {Jarillo-Herrero}}]{Ma2018}%
  \BibitemOpen
  \bibfield  {author} {\bibinfo {author} {\bibfnamefont {Q.}~\bibnamefont
  {Ma}}, \bibinfo {author} {\bibfnamefont {S.-Y.}\ \bibnamefont {Xu}}, \bibinfo
  {author} {\bibfnamefont {H.}~\bibnamefont {Shen}}, \bibinfo {author}
  {\bibfnamefont {D.}~\bibnamefont {MacNeill}}, \bibinfo {author}
  {\bibfnamefont {V.}~\bibnamefont {Fatemi}}, \bibinfo {author} {\bibfnamefont
  {T.-R.}\ \bibnamefont {Chang}}, \bibinfo {author} {\bibfnamefont {A.~M.~M.}\
  \bibnamefont {Valdivia}}, \bibinfo {author} {\bibfnamefont {S.}~\bibnamefont
  {Wu}}, \bibinfo {author} {\bibfnamefont {Z.}~\bibnamefont {Du}}, \bibinfo
  {author} {\bibfnamefont {C.-H.}\ \bibnamefont {Hsu}}, \bibinfo {author}
  {\bibfnamefont {S.}~\bibnamefont {Fang}}, \bibinfo {author} {\bibfnamefont
  {Q.~D.}\ \bibnamefont {Gibson}}, \bibinfo {author} {\bibfnamefont
  {K.}~\bibnamefont {Watanabe}}, \bibinfo {author} {\bibfnamefont
  {T.}~\bibnamefont {Taniguchi}}, \bibinfo {author} {\bibfnamefont {R.~J.}\
  \bibnamefont {Cava}}, \bibinfo {author} {\bibfnamefont {E.}~\bibnamefont
  {Kaxiras}}, \bibinfo {author} {\bibfnamefont {H.-Z.}\ \bibnamefont {Lu}},
  \bibinfo {author} {\bibfnamefont {H.}~\bibnamefont {Lin}}, \bibinfo {author}
  {\bibfnamefont {L.}~\bibnamefont {Fu}}, \bibinfo {author} {\bibfnamefont
  {N.}~\bibnamefont {Gedik}}, \ and\ \bibinfo {author} {\bibfnamefont
  {P.}~\bibnamefont {Jarillo-Herrero}},\ }\href {\doibase
  10.1038/s41586-018-0807-6} {\bibfield  {journal} {\bibinfo  {journal}
  {Nature}\ }\textbf {\bibinfo {volume} {565}},\ \bibinfo {pages} {337}
  (\bibinfo {year} {2018})}\BibitemShut {NoStop}%
\bibitem [{\citenamefont {Zhang}\ \emph
  {et~al.}(2018{\natexlab{b}})\citenamefont {Zhang}, \citenamefont {van~den
  Brink}, \citenamefont {Felser},\ and\ \citenamefont {Yan}}]{Zhang2018I}%
  \BibitemOpen
  \bibfield  {author} {\bibinfo {author} {\bibfnamefont {Y.}~\bibnamefont
  {Zhang}}, \bibinfo {author} {\bibfnamefont {J.}~\bibnamefont {van~den
  Brink}}, \bibinfo {author} {\bibfnamefont {C.}~\bibnamefont {Felser}}, \ and\
  \bibinfo {author} {\bibfnamefont {B.}~\bibnamefont {Yan}},\ }\href {\doibase
  10.1088/2053-1583/aad1ae} {\bibfield  {journal} {\bibinfo  {journal} {2D
  Materials}\ }\textbf {\bibinfo {volume} {5}},\ \bibinfo {pages} {044001}
  (\bibinfo {year} {2018}{\natexlab{b}})}\BibitemShut {NoStop}%
\bibitem [{\citenamefont {Du}\ \emph {et~al.}(2019)\citenamefont {Du},
  \citenamefont {Wang}, \citenamefont {Li}, \citenamefont {Lu},\ and\
  \citenamefont {Xie}}]{Du2019}%
  \BibitemOpen
  \bibfield  {author} {\bibinfo {author} {\bibfnamefont {Z.~Z.}\ \bibnamefont
  {Du}}, \bibinfo {author} {\bibfnamefont {C.~M.}\ \bibnamefont {Wang}},
  \bibinfo {author} {\bibfnamefont {S.}~\bibnamefont {Li}}, \bibinfo {author}
  {\bibfnamefont {H.-Z.}\ \bibnamefont {Lu}}, \ and\ \bibinfo {author}
  {\bibfnamefont {X.~C.}\ \bibnamefont {Xie}},\ }\href {\doibase
  10.1038/s41467-019-10941-3} {\bibfield  {journal} {\bibinfo  {journal}
  {Nature Communications}\ }\textbf {\bibinfo {volume} {10}} (\bibinfo {year}
  {2019}),\ 10.1038/s41467-019-10941-3}\BibitemShut {NoStop}%
\bibitem [{\citenamefont {K\"onig}\ \emph {et~al.}(2019)\citenamefont
  {K\"onig}, \citenamefont {Dzero}, \citenamefont {Levchenko},\ and\
  \citenamefont {Pesin}}]{Konig2019}%
  \BibitemOpen
  \bibfield  {author} {\bibinfo {author} {\bibfnamefont {E.~J.}\ \bibnamefont
  {K\"onig}}, \bibinfo {author} {\bibfnamefont {M.}~\bibnamefont {Dzero}},
  \bibinfo {author} {\bibfnamefont {A.}~\bibnamefont {Levchenko}}, \ and\
  \bibinfo {author} {\bibfnamefont {D.~A.}\ \bibnamefont {Pesin}},\ }\href
  {\doibase 10.1103/PhysRevB.99.155404} {\bibfield  {journal} {\bibinfo
  {journal} {Phys. Rev. B}\ }\textbf {\bibinfo {volume} {99}},\ \bibinfo
  {pages} {155404} (\bibinfo {year} {2019})}\BibitemShut {NoStop}%
\bibitem [{\citenamefont {Kang}\ \emph {et~al.}(2019)\citenamefont {Kang},
  \citenamefont {Li}, \citenamefont {Sohn}, \citenamefont {Shan},\ and\
  \citenamefont {Mak}}]{Kang2019}%
  \BibitemOpen
  \bibfield  {author} {\bibinfo {author} {\bibfnamefont {K.}~\bibnamefont
  {Kang}}, \bibinfo {author} {\bibfnamefont {T.}~\bibnamefont {Li}}, \bibinfo
  {author} {\bibfnamefont {E.}~\bibnamefont {Sohn}}, \bibinfo {author}
  {\bibfnamefont {J.}~\bibnamefont {Shan}}, \ and\ \bibinfo {author}
  {\bibfnamefont {K.~F.}\ \bibnamefont {Mak}},\ }\href {\doibase
  10.1038/s41563-019-0294-7} {\bibfield  {journal} {\bibinfo  {journal} {Nature
  Materials}\ }\textbf {\bibinfo {volume} {18}},\ \bibinfo {pages} {324}
  (\bibinfo {year} {2019})}\BibitemShut {NoStop}%
\bibitem [{\citenamefont {Son}\ \emph {et~al.}(2019)\citenamefont {Son},
  \citenamefont {Kim}, \citenamefont {Ahn}, \citenamefont {Lee},\ and\
  \citenamefont {Lee}}]{Son2019}%
  \BibitemOpen
  \bibfield  {author} {\bibinfo {author} {\bibfnamefont {J.}~\bibnamefont
  {Son}}, \bibinfo {author} {\bibfnamefont {K.-H.}\ \bibnamefont {Kim}},
  \bibinfo {author} {\bibfnamefont {Y.~H.}\ \bibnamefont {Ahn}}, \bibinfo
  {author} {\bibfnamefont {H.-W.}\ \bibnamefont {Lee}}, \ and\ \bibinfo
  {author} {\bibfnamefont {J.}~\bibnamefont {Lee}},\ }\href {\doibase
  10.1103/PhysRevLett.123.036806} {\bibfield  {journal} {\bibinfo  {journal}
  {Phys. Rev. Lett.}\ }\textbf {\bibinfo {volume} {123}},\ \bibinfo {pages}
  {036806} (\bibinfo {year} {2019})}\BibitemShut {NoStop}%
\bibitem [{\citenamefont {Battilomo}\ \emph {et~al.}(2019)\citenamefont
  {Battilomo}, \citenamefont {Scopigno},\ and\ \citenamefont
  {Ortix}}]{Battilomo2019}%
  \BibitemOpen
  \bibfield  {author} {\bibinfo {author} {\bibfnamefont {R.}~\bibnamefont
  {Battilomo}}, \bibinfo {author} {\bibfnamefont {N.}~\bibnamefont {Scopigno}},
  \ and\ \bibinfo {author} {\bibfnamefont {C.}~\bibnamefont {Ortix}},\ }\href
  {\doibase 10.1103/PhysRevLett.123.196403} {\bibfield  {journal} {\bibinfo
  {journal} {Phys. Rev. Lett.}\ }\textbf {\bibinfo {volume} {123}},\ \bibinfo
  {pages} {196403} (\bibinfo {year} {2019})}\BibitemShut {NoStop}%
\bibitem [{\citenamefont {Zeng}\ \emph {et~al.}(2019)\citenamefont {Zeng},
  \citenamefont {Nandy}, \citenamefont {Taraphder},\ and\ \citenamefont
  {Tewari}}]{Zeng2019}%
  \BibitemOpen
  \bibfield  {author} {\bibinfo {author} {\bibfnamefont {C.}~\bibnamefont
  {Zeng}}, \bibinfo {author} {\bibfnamefont {S.}~\bibnamefont {Nandy}},
  \bibinfo {author} {\bibfnamefont {A.}~\bibnamefont {Taraphder}}, \ and\
  \bibinfo {author} {\bibfnamefont {S.}~\bibnamefont {Tewari}},\ }\href
  {\doibase 10.1103/PhysRevB.100.245102} {\bibfield  {journal} {\bibinfo
  {journal} {Phys. Rev. B}\ }\textbf {\bibinfo {volume} {100}},\ \bibinfo
  {pages} {245102} (\bibinfo {year} {2019})}\BibitemShut {NoStop}%
\bibitem [{\citenamefont {Yu}\ \emph {et~al.}(2019)\citenamefont {Yu},
  \citenamefont {Zhu}, \citenamefont {You}, \citenamefont {Low},\ and\
  \citenamefont {Su}}]{Yu2019}%
  \BibitemOpen
  \bibfield  {author} {\bibinfo {author} {\bibfnamefont {X.-Q.}\ \bibnamefont
  {Yu}}, \bibinfo {author} {\bibfnamefont {Z.-G.}\ \bibnamefont {Zhu}},
  \bibinfo {author} {\bibfnamefont {J.-S.}\ \bibnamefont {You}}, \bibinfo
  {author} {\bibfnamefont {T.}~\bibnamefont {Low}}, \ and\ \bibinfo {author}
  {\bibfnamefont {G.}~\bibnamefont {Su}},\ }\href {\doibase
  10.1103/PhysRevB.99.201410} {\bibfield  {journal} {\bibinfo  {journal} {Phys.
  Rev. B}\ }\textbf {\bibinfo {volume} {99}},\ \bibinfo {pages} {201410}
  (\bibinfo {year} {2019})}\BibitemShut {NoStop}%
\bibitem [{\citenamefont {Kumar}\ \emph {et~al.}(2021)\citenamefont {Kumar},
  \citenamefont {Hsu}, \citenamefont {Sharma}, \citenamefont {Chang},
  \citenamefont {Yu}, \citenamefont {Wang}, \citenamefont {Eda}, \citenamefont
  {Liang},\ and\ \citenamefont {Yang}}]{Kumar2021}%
  \BibitemOpen
  \bibfield  {author} {\bibinfo {author} {\bibfnamefont {D.}~\bibnamefont
  {Kumar}}, \bibinfo {author} {\bibfnamefont {C.-H.}\ \bibnamefont {Hsu}},
  \bibinfo {author} {\bibfnamefont {R.}~\bibnamefont {Sharma}}, \bibinfo
  {author} {\bibfnamefont {T.-R.}\ \bibnamefont {Chang}}, \bibinfo {author}
  {\bibfnamefont {P.}~\bibnamefont {Yu}}, \bibinfo {author} {\bibfnamefont
  {J.}~\bibnamefont {Wang}}, \bibinfo {author} {\bibfnamefont {G.}~\bibnamefont
  {Eda}}, \bibinfo {author} {\bibfnamefont {G.}~\bibnamefont {Liang}}, \ and\
  \bibinfo {author} {\bibfnamefont {H.}~\bibnamefont {Yang}},\ }\href {\doibase
  10.1038/s41565-020-00839-3} {\bibfield  {journal} {\bibinfo  {journal}
  {Nature Nanotechnology}\ }\textbf {\bibinfo {volume} {16}},\ \bibinfo {pages}
  {421} (\bibinfo {year} {2021})}\BibitemShut {NoStop}%
\bibitem [{\citenamefont {Ortix}(2021)}]{Ortix2021}%
  \BibitemOpen
  \bibfield  {author} {\bibinfo {author} {\bibfnamefont {C.}~\bibnamefont
  {Ortix}},\ }\href {\doibase 10.1002/qute.202100056} {\bibfield  {journal}
  {\bibinfo  {journal} {Advanced Quantum Technologies}\ }\textbf {\bibinfo
  {volume} {4}},\ \bibinfo {pages} {2100056} (\bibinfo {year}
  {2021})}\BibitemShut {NoStop}%
\bibitem [{\citenamefont {Du}\ \emph {et~al.}(2021)\citenamefont {Du},
  \citenamefont {Wang}, \citenamefont {Sun}, \citenamefont {Lu},\ and\
  \citenamefont {Xie}}]{Du2021}%
  \BibitemOpen
  \bibfield  {author} {\bibinfo {author} {\bibfnamefont {Z.~Z.}\ \bibnamefont
  {Du}}, \bibinfo {author} {\bibfnamefont {C.~M.}\ \bibnamefont {Wang}},
  \bibinfo {author} {\bibfnamefont {H.-P.}\ \bibnamefont {Sun}}, \bibinfo
  {author} {\bibfnamefont {H.-Z.}\ \bibnamefont {Lu}}, \ and\ \bibinfo {author}
  {\bibfnamefont {X.~C.}\ \bibnamefont {Xie}},\ }\href {\doibase
  10.1038/s41467-021-25273-4} {\bibfield  {journal} {\bibinfo  {journal}
  {Nature Communications}\ }\textbf {\bibinfo {volume} {12}} (\bibinfo {year}
  {2021}),\ 10.1038/s41467-021-25273-4}\BibitemShut {NoStop}%
\bibitem [{\citenamefont {Lai}\ \emph {et~al.}(2021)\citenamefont {Lai},
  \citenamefont {Liu}, \citenamefont {Zhang}, \citenamefont {Zhao},
  \citenamefont {Feng}, \citenamefont {Wang}, \citenamefont {Tang},
  \citenamefont {Liu}, \citenamefont {Novoselov}, \citenamefont {Yang},\ and\
  \citenamefont {bo~Gao}}]{Lai2021}%
  \BibitemOpen
  \bibfield  {author} {\bibinfo {author} {\bibfnamefont {S.}~\bibnamefont
  {Lai}}, \bibinfo {author} {\bibfnamefont {H.}~\bibnamefont {Liu}}, \bibinfo
  {author} {\bibfnamefont {Z.}~\bibnamefont {Zhang}}, \bibinfo {author}
  {\bibfnamefont {J.}~\bibnamefont {Zhao}}, \bibinfo {author} {\bibfnamefont
  {X.}~\bibnamefont {Feng}}, \bibinfo {author} {\bibfnamefont {N.}~\bibnamefont
  {Wang}}, \bibinfo {author} {\bibfnamefont {C.}~\bibnamefont {Tang}}, \bibinfo
  {author} {\bibfnamefont {Y.}~\bibnamefont {Liu}}, \bibinfo {author}
  {\bibfnamefont {K.~S.}\ \bibnamefont {Novoselov}}, \bibinfo {author}
  {\bibfnamefont {S.~A.}\ \bibnamefont {Yang}}, \ and\ \bibinfo {author}
  {\bibfnamefont {W.}~\bibnamefont {bo~Gao}},\ }\href {\doibase
  10.1038/s41565-021-00917-0} {\bibfield  {journal} {\bibinfo  {journal} {Nat.
  Nanotechnol}\ }\textbf {\bibinfo {volume} {16}},\ \bibinfo {pages} {869}
  (\bibinfo {year} {2021})}\BibitemShut {NoStop}%
\bibitem [{\citenamefont {Ye}\ \emph {et~al.}(2022)\citenamefont {Ye},
  \citenamefont {Zhu}, \citenamefont {Xu}, \citenamefont {Zang}, \citenamefont
  {Ye},\ and\ \citenamefont {Liao}}]{Min_2022}%
  \BibitemOpen
  \bibfield  {author} {\bibinfo {author} {\bibfnamefont {X.-G.}\ \bibnamefont
  {Ye}}, \bibinfo {author} {\bibfnamefont {P.-F.}\ \bibnamefont {Zhu}},
  \bibinfo {author} {\bibfnamefont {W.-Z.}\ \bibnamefont {Xu}}, \bibinfo
  {author} {\bibfnamefont {Z.}~\bibnamefont {Zang}}, \bibinfo {author}
  {\bibfnamefont {Y.}~\bibnamefont {Ye}}, \ and\ \bibinfo {author}
  {\bibfnamefont {Z.-M.}\ \bibnamefont {Liao}},\ }\href {\doibase
  10.1103/PhysRevB.106.045414} {\bibfield  {journal} {\bibinfo  {journal}
  {Phys. Rev. B}\ }\textbf {\bibinfo {volume} {106}},\ \bibinfo {pages}
  {045414} (\bibinfo {year} {2022})}\BibitemShut {NoStop}%
\bibitem [{\citenamefont {Ma}\ \emph {et~al.}(2021)\citenamefont {Ma},
  \citenamefont {Muniz}, \citenamefont {Qi}, \citenamefont {Lai}, \citenamefont
  {Zhang}, \citenamefont {Liu}, \citenamefont {Zhuo}, \citenamefont {Chen},
  \citenamefont {Chen}, \citenamefont {Zhou},\ and\ \citenamefont
  {Sun}}]{Ma2021}%
  \BibitemOpen
  \bibfield  {author} {\bibinfo {author} {\bibfnamefont {J.}~\bibnamefont
  {Ma}}, \bibinfo {author} {\bibfnamefont {R.~A.}\ \bibnamefont {Muniz}},
  \bibinfo {author} {\bibfnamefont {S.}~\bibnamefont {Qi}}, \bibinfo {author}
  {\bibfnamefont {J.}~\bibnamefont {Lai}}, \bibinfo {author} {\bibfnamefont
  {K.}~\bibnamefont {Zhang}}, \bibinfo {author} {\bibfnamefont
  {Y.}~\bibnamefont {Liu}}, \bibinfo {author} {\bibfnamefont {X.}~\bibnamefont
  {Zhuo}}, \bibinfo {author} {\bibfnamefont {S.}~\bibnamefont {Chen}}, \bibinfo
  {author} {\bibfnamefont {J.-H.}\ \bibnamefont {Chen}}, \bibinfo {author}
  {\bibfnamefont {S.}~\bibnamefont {Zhou}}, \ and\ \bibinfo {author}
  {\bibfnamefont {D.}~\bibnamefont {Sun}},\ }\href {\doibase
  10.1088/2053-1583/abd6b3} {\bibfield  {journal} {\bibinfo  {journal} {2D
  Materials}\ }\textbf {\bibinfo {volume} {8}},\ \bibinfo {pages} {025016}
  (\bibinfo {year} {2021})}\BibitemShut {NoStop}%
\bibitem [{\citenamefont {Gao}\ \emph {et~al.}(2014)\citenamefont {Gao},
  \citenamefont {Yang},\ and\ \citenamefont {Niu}}]{Gao2014}%
  \BibitemOpen
  \bibfield  {author} {\bibinfo {author} {\bibfnamefont {Y.}~\bibnamefont
  {Gao}}, \bibinfo {author} {\bibfnamefont {S.~A.}\ \bibnamefont {Yang}}, \
  and\ \bibinfo {author} {\bibfnamefont {Q.}~\bibnamefont {Niu}},\ }\href
  {\doibase 10.1103/PhysRevLett.112.166601} {\bibfield  {journal} {\bibinfo
  {journal} {Phys. Rev. Lett.}\ }\textbf {\bibinfo {volume} {112}},\ \bibinfo
  {pages} {166601} (\bibinfo {year} {2014})}\BibitemShut {NoStop}%
\bibitem [{\citenamefont {Ashcroft}\ and\ \citenamefont
  {Mermin}(1976)}]{ashcroft1976solid}%
  \BibitemOpen
  \bibfield  {author} {\bibinfo {author} {\bibfnamefont {N.~W.}\ \bibnamefont
  {Ashcroft}}\ and\ \bibinfo {author} {\bibfnamefont {N.~D.}\ \bibnamefont
  {Mermin}},\ }\href@noop {} {\emph {\bibinfo {title} {Solid state physics}}}\
  (\bibinfo  {publisher} {New York: Holt, Rinehart and Winston,},\ \bibinfo
  {year} {1976})\BibitemShut {NoStop}%
\bibitem [{\citenamefont {Ziman}(2001)}]{ziman2001electrons}%
  \BibitemOpen
  \bibfield  {author} {\bibinfo {author} {\bibfnamefont {J.~M.}\ \bibnamefont
  {Ziman}},\ }\href@noop {} {\emph {\bibinfo {title} {Electrons and phonons:
  the theory of transport phenomena in solids}}}\ (\bibinfo  {publisher}
  {Oxford university press},\ \bibinfo {year} {2001})\BibitemShut {NoStop}%
\bibitem [{\citenamefont {Nandy}\ \emph {et~al.}(2017)\citenamefont {Nandy},
  \citenamefont {Sharma}, \citenamefont {Taraphder},\ and\ \citenamefont
  {Tewari}}]{Nandy_2017}%
  \BibitemOpen
  \bibfield  {author} {\bibinfo {author} {\bibfnamefont {S.}~\bibnamefont
  {Nandy}}, \bibinfo {author} {\bibfnamefont {G.}~\bibnamefont {Sharma}},
  \bibinfo {author} {\bibfnamefont {A.}~\bibnamefont {Taraphder}}, \ and\
  \bibinfo {author} {\bibfnamefont {S.}~\bibnamefont {Tewari}},\ }\href
  {\doibase 10.1103/PhysRevLett.119.176804} {\bibfield  {journal} {\bibinfo
  {journal} {Phys. Rev. Lett.}\ }\textbf {\bibinfo {volume} {119}},\ \bibinfo
  {pages} {176804} (\bibinfo {year} {2017})}\BibitemShut {NoStop}%
\bibitem [{\citenamefont {Nag}\ and\ \citenamefont {Nandy}(2020)}]{Nag_2020}%
  \BibitemOpen
  \bibfield  {author} {\bibinfo {author} {\bibfnamefont {T.}~\bibnamefont
  {Nag}}\ and\ \bibinfo {author} {\bibfnamefont {S.}~\bibnamefont {Nandy}},\
  }\href {\doibase 10.1088/1361-648x/abc310} {\bibfield  {journal} {\bibinfo
  {journal} {Journal of Physics: Condensed Matter}\ }\textbf {\bibinfo {volume}
  {33}},\ \bibinfo {pages} {075504} (\bibinfo {year} {2020})}\BibitemShut
  {NoStop}%
\bibitem [{\citenamefont {Xiang}\ \emph {et~al.}(2023)\citenamefont {Xiang},
  \citenamefont {Zhang}, \citenamefont {Wang},\ and\ \citenamefont
  {Wang}}]{xiang2023third}%
  \BibitemOpen
  \bibfield  {author} {\bibinfo {author} {\bibfnamefont {L.}~\bibnamefont
  {Xiang}}, \bibinfo {author} {\bibfnamefont {C.}~\bibnamefont {Zhang}},
  \bibinfo {author} {\bibfnamefont {L.}~\bibnamefont {Wang}}, \ and\ \bibinfo
  {author} {\bibfnamefont {J.}~\bibnamefont {Wang}},\ }\href@noop {} {\bibfield
   {journal} {\bibinfo  {journal} {Physical Review B}\ }\textbf {\bibinfo
  {volume} {107}},\ \bibinfo {pages} {075411} (\bibinfo {year}
  {2023})}\BibitemShut {NoStop}%
\bibitem [{\citenamefont {He}\ \emph {et~al.}(2019)\citenamefont {He},
  \citenamefont {Zhang}, \citenamefont {Zhu}, \citenamefont {Shi},
  \citenamefont {Heinonen}, \citenamefont {Vignale},\ and\ \citenamefont
  {Yang}}]{Steven19}%
  \BibitemOpen
  \bibfield  {author} {\bibinfo {author} {\bibfnamefont {P.}~\bibnamefont
  {He}}, \bibinfo {author} {\bibfnamefont {S.~S.-L.}\ \bibnamefont {Zhang}},
  \bibinfo {author} {\bibfnamefont {D.}~\bibnamefont {Zhu}}, \bibinfo {author}
  {\bibfnamefont {S.}~\bibnamefont {Shi}}, \bibinfo {author} {\bibfnamefont
  {O.~G.}\ \bibnamefont {Heinonen}}, \bibinfo {author} {\bibfnamefont
  {G.}~\bibnamefont {Vignale}}, \ and\ \bibinfo {author} {\bibfnamefont
  {H.}~\bibnamefont {Yang}},\ }\href {\doibase 10.1103/PhysRevLett.123.016801}
  {\bibfield  {journal} {\bibinfo  {journal} {Phys. Rev. Lett.}\ }\textbf
  {\bibinfo {volume} {123}},\ \bibinfo {pages} {016801} (\bibinfo {year}
  {2019})}\BibitemShut {NoStop}%
\bibitem [{\citenamefont {Rao}\ \emph {et~al.}(2021)\citenamefont {Rao},
  \citenamefont {Zhou}, \citenamefont {Wu}, \citenamefont {Duan}, \citenamefont
  {Deng},\ and\ \citenamefont {Wang}}]{Zhou21}%
  \BibitemOpen
  \bibfield  {author} {\bibinfo {author} {\bibfnamefont {W.}~\bibnamefont
  {Rao}}, \bibinfo {author} {\bibfnamefont {Y.-L.}\ \bibnamefont {Zhou}},
  \bibinfo {author} {\bibfnamefont {Y.-j.}\ \bibnamefont {Wu}}, \bibinfo
  {author} {\bibfnamefont {H.-J.}\ \bibnamefont {Duan}}, \bibinfo {author}
  {\bibfnamefont {M.-X.}\ \bibnamefont {Deng}}, \ and\ \bibinfo {author}
  {\bibfnamefont {R.-Q.}\ \bibnamefont {Wang}},\ }\href {\doibase
  10.1103/PhysRevB.103.155415} {\bibfield  {journal} {\bibinfo  {journal}
  {Phys. Rev. B}\ }\textbf {\bibinfo {volume} {103}},\ \bibinfo {pages}
  {155415} (\bibinfo {year} {2021})}\BibitemShut {NoStop}%
\bibitem [{\citenamefont {Battilomo}\ \emph {et~al.}(2021)\citenamefont
  {Battilomo}, \citenamefont {Scopigno},\ and\ \citenamefont
  {Ortix}}]{Battilomo21}%
  \BibitemOpen
  \bibfield  {author} {\bibinfo {author} {\bibfnamefont {R.}~\bibnamefont
  {Battilomo}}, \bibinfo {author} {\bibfnamefont {N.}~\bibnamefont {Scopigno}},
  \ and\ \bibinfo {author} {\bibfnamefont {C.}~\bibnamefont {Ortix}},\ }\href
  {\doibase 10.1103/PhysRevResearch.3.L012006} {\bibfield  {journal} {\bibinfo
  {journal} {Phys. Rev. Research}\ }\textbf {\bibinfo {volume} {3}},\ \bibinfo
  {pages} {L012006} (\bibinfo {year} {2021})}\BibitemShut {NoStop}%
\bibitem [{\citenamefont {Huang}\ \emph {et~al.}(2022)\citenamefont {Huang},
  \citenamefont {Feng}, \citenamefont {Wang}, \citenamefont {Xiao},\ and\
  \citenamefont {Yang}}]{huang2022intrinsic}%
  \BibitemOpen
  \bibfield  {author} {\bibinfo {author} {\bibfnamefont {Y.-X.}\ \bibnamefont
  {Huang}}, \bibinfo {author} {\bibfnamefont {X.}~\bibnamefont {Feng}},
  \bibinfo {author} {\bibfnamefont {H.}~\bibnamefont {Wang}}, \bibinfo {author}
  {\bibfnamefont {C.}~\bibnamefont {Xiao}}, \ and\ \bibinfo {author}
  {\bibfnamefont {S.~A.}\ \bibnamefont {Yang}},\ }\href {\doibase
  10.48550/ARXIV.2208.03639} {\enquote {\bibinfo {title} {Intrinsic nonlinear
  planar hall effect},}\ } (\bibinfo {year} {2022})\BibitemShut {NoStop}%
\bibitem [{\citenamefont {Fu}(2009)}]{Fu09}%
  \BibitemOpen
  \bibfield  {author} {\bibinfo {author} {\bibfnamefont {L.}~\bibnamefont
  {Fu}},\ }\href {\doibase 10.1103/PhysRevLett.103.266801} {\bibfield
  {journal} {\bibinfo  {journal} {Phys. Rev. Lett.}\ }\textbf {\bibinfo
  {volume} {103}},\ \bibinfo {pages} {266801} (\bibinfo {year}
  {2009})}\BibitemShut {NoStop}%
\bibitem [{\citenamefont {Zhang}\ \emph {et~al.}(2012)\citenamefont {Zhang},
  \citenamefont {Kane},\ and\ \citenamefont {Mele}}]{tilt_TSS_2012_zhang}%
  \BibitemOpen
  \bibfield  {author} {\bibinfo {author} {\bibfnamefont {F.}~\bibnamefont
  {Zhang}}, \bibinfo {author} {\bibfnamefont {C.~L.}\ \bibnamefont {Kane}}, \
  and\ \bibinfo {author} {\bibfnamefont {E.~J.}\ \bibnamefont {Mele}},\ }\href
  {\doibase 10.1103/PhysRevB.86.081303} {\bibfield  {journal} {\bibinfo
  {journal} {Phys. Rev. B}\ }\textbf {\bibinfo {volume} {86}},\ \bibinfo
  {pages} {081303} (\bibinfo {year} {2012})}\BibitemShut {NoStop}%
\bibitem [{\citenamefont {Zhang}\ \emph {et~al.}(2013)\citenamefont {Zhang},
  \citenamefont {Kane},\ and\ \citenamefont
  {Mele}}]{breakingTRS_Zhang_2013PRL}%
  \BibitemOpen
  \bibfield  {author} {\bibinfo {author} {\bibfnamefont {F.}~\bibnamefont
  {Zhang}}, \bibinfo {author} {\bibfnamefont {C.~L.}\ \bibnamefont {Kane}}, \
  and\ \bibinfo {author} {\bibfnamefont {E.~J.}\ \bibnamefont {Mele}},\ }\href
  {\doibase 10.1103/PhysRevLett.110.046404} {\bibfield  {journal} {\bibinfo
  {journal} {Phys. Rev. Lett.}\ }\textbf {\bibinfo {volume} {110}},\ \bibinfo
  {pages} {046404} (\bibinfo {year} {2013})}\BibitemShut {NoStop}%
\bibitem [{\citenamefont {Somroob}\ \emph {et~al.}(2021)\citenamefont
  {Somroob}, \citenamefont {Sutthibutpong}, \citenamefont {Tangwancharoen},\
  and\ \citenamefont {Liewrian}}]{somroob2021tunable}%
  \BibitemOpen
  \bibfield  {author} {\bibinfo {author} {\bibfnamefont {P.}~\bibnamefont
  {Somroob}}, \bibinfo {author} {\bibfnamefont {T.}~\bibnamefont
  {Sutthibutpong}}, \bibinfo {author} {\bibfnamefont {S.}~\bibnamefont
  {Tangwancharoen}}, \ and\ \bibinfo {author} {\bibfnamefont {W.}~\bibnamefont
  {Liewrian}},\ }\href@noop {} {\bibfield  {journal} {\bibinfo  {journal}
  {Physica E: Low-dimensional Systems and Nanostructures}\ }\textbf {\bibinfo
  {volume} {127}},\ \bibinfo {pages} {114501} (\bibinfo {year}
  {2021})}\BibitemShut {NoStop}%
\bibitem [{\citenamefont {Zheng}\ \emph {et~al.}(2020)\citenamefont {Zheng},
  \citenamefont {Duan}, \citenamefont {Wang}, \citenamefont {Li}, \citenamefont
  {Deng},\ and\ \citenamefont {Wang}}]{PHE_tilt_2020}%
  \BibitemOpen
  \bibfield  {author} {\bibinfo {author} {\bibfnamefont {S.-H.}\ \bibnamefont
  {Zheng}}, \bibinfo {author} {\bibfnamefont {H.-J.}\ \bibnamefont {Duan}},
  \bibinfo {author} {\bibfnamefont {J.-K.}\ \bibnamefont {Wang}}, \bibinfo
  {author} {\bibfnamefont {J.-Y.}\ \bibnamefont {Li}}, \bibinfo {author}
  {\bibfnamefont {M.-X.}\ \bibnamefont {Deng}}, \ and\ \bibinfo {author}
  {\bibfnamefont {R.-Q.}\ \bibnamefont {Wang}},\ }\href {\doibase
  10.1103/PhysRevB.101.041408} {\bibfield  {journal} {\bibinfo  {journal}
  {Phys. Rev. B}\ }\textbf {\bibinfo {volume} {101}},\ \bibinfo {pages}
  {041408} (\bibinfo {year} {2020})}\BibitemShut {NoStop}%
\bibitem [{\citenamefont {Zhang}\ \emph {et~al.}(2021)\citenamefont {Zhang},
  \citenamefont {Li}, \citenamefont {Pe\~na Benitez}, \citenamefont
  {Sur\'owka}, \citenamefont {Moessner}, \citenamefont {Molenkamp},\ and\
  \citenamefont {Trauzettel}}]{super_transport_2021}%
  \BibitemOpen
  \bibfield  {author} {\bibinfo {author} {\bibfnamefont {S.-B.}\ \bibnamefont
  {Zhang}}, \bibinfo {author} {\bibfnamefont {C.-A.}\ \bibnamefont {Li}},
  \bibinfo {author} {\bibfnamefont {F.}~\bibnamefont {Pe\~na Benitez}},
  \bibinfo {author} {\bibfnamefont {P.}~\bibnamefont {Sur\'owka}}, \bibinfo
  {author} {\bibfnamefont {R.}~\bibnamefont {Moessner}}, \bibinfo {author}
  {\bibfnamefont {L.~W.}\ \bibnamefont {Molenkamp}}, \ and\ \bibinfo {author}
  {\bibfnamefont {B.}~\bibnamefont {Trauzettel}},\ }\href {\doibase
  10.1103/PhysRevLett.127.076601} {\bibfield  {journal} {\bibinfo  {journal}
  {Phys. Rev. Lett.}\ }\textbf {\bibinfo {volume} {127}},\ \bibinfo {pages}
  {076601} (\bibinfo {year} {2021})}\BibitemShut {NoStop}%
\bibitem [{\citenamefont {Zeng}\ \emph {et~al.}(2022)\citenamefont {Zeng},
  \citenamefont {Yu}, \citenamefont {Yu},\ and\ \citenamefont
  {Yao}}]{NLNE_czeng_2022}%
  \BibitemOpen
  \bibfield  {author} {\bibinfo {author} {\bibfnamefont {C.}~\bibnamefont
  {Zeng}}, \bibinfo {author} {\bibfnamefont {X.-Q.}\ \bibnamefont {Yu}},
  \bibinfo {author} {\bibfnamefont {Z.-M.}\ \bibnamefont {Yu}}, \ and\ \bibinfo
  {author} {\bibfnamefont {Y.}~\bibnamefont {Yao}},\ }\href {\doibase
  10.1103/PhysRevB.106.L081121} {\bibfield  {journal} {\bibinfo  {journal}
  {Phys. Rev. B}\ }\textbf {\bibinfo {volume} {106}},\ \bibinfo {pages}
  {L081121} (\bibinfo {year} {2022})}\BibitemShut {NoStop}%
\bibitem [{\citenamefont {He}\ \emph {et~al.}(2018)\citenamefont {He},
  \citenamefont {Zhang}, \citenamefont {Zhu}, \citenamefont {Liu},
  \citenamefont {Wang}, \citenamefont {Yu}, \citenamefont {Vignale},\ and\
  \citenamefont {Yang}}]{He_2018}%
  \BibitemOpen
  \bibfield  {author} {\bibinfo {author} {\bibfnamefont {P.}~\bibnamefont
  {He}}, \bibinfo {author} {\bibfnamefont {S.~S.-L.}\ \bibnamefont {Zhang}},
  \bibinfo {author} {\bibfnamefont {D.}~\bibnamefont {Zhu}}, \bibinfo {author}
  {\bibfnamefont {Y.}~\bibnamefont {Liu}}, \bibinfo {author} {\bibfnamefont
  {Y.}~\bibnamefont {Wang}}, \bibinfo {author} {\bibfnamefont {J.}~\bibnamefont
  {Yu}}, \bibinfo {author} {\bibfnamefont {G.}~\bibnamefont {Vignale}}, \ and\
  \bibinfo {author} {\bibfnamefont {H.}~\bibnamefont {Yang}},\ }\href {\doibase
  10.1038/s41567-017-0039-y} {\bibfield  {journal} {\bibinfo  {journal} {Nature
  Physics}\ }\textbf {\bibinfo {volume} {14}},\ \bibinfo {pages} {495}
  (\bibinfo {year} {2018})}\BibitemShut {NoStop}%
\bibitem [{\citenamefont {Alpichshev}\ \emph {et~al.}(2010)\citenamefont
  {Alpichshev}, \citenamefont {Analytis}, \citenamefont {Chu}, \citenamefont
  {Fisher}, \citenamefont {Chen}, \citenamefont {Shen}, \citenamefont {Fang},\
  and\ \citenamefont {Kapitulnik}}]{tss_2010prl}%
  \BibitemOpen
  \bibfield  {author} {\bibinfo {author} {\bibfnamefont {Z.}~\bibnamefont
  {Alpichshev}}, \bibinfo {author} {\bibfnamefont {J.~G.}\ \bibnamefont
  {Analytis}}, \bibinfo {author} {\bibfnamefont {J.-H.}\ \bibnamefont {Chu}},
  \bibinfo {author} {\bibfnamefont {I.~R.}\ \bibnamefont {Fisher}}, \bibinfo
  {author} {\bibfnamefont {Y.~L.}\ \bibnamefont {Chen}}, \bibinfo {author}
  {\bibfnamefont {Z.~X.}\ \bibnamefont {Shen}}, \bibinfo {author}
  {\bibfnamefont {A.}~\bibnamefont {Fang}}, \ and\ \bibinfo {author}
  {\bibfnamefont {A.}~\bibnamefont {Kapitulnik}},\ }\href {\doibase
  10.1103/PhysRevLett.104.016401} {\bibfield  {journal} {\bibinfo  {journal}
  {Phys. Rev. Lett.}\ }\textbf {\bibinfo {volume} {104}},\ \bibinfo {pages}
  {016401} (\bibinfo {year} {2010})}\BibitemShut {NoStop}%
\bibitem [{\citenamefont {Nomura}\ \emph {et~al.}(2014)\citenamefont {Nomura},
  \citenamefont {Souma}, \citenamefont {Takayama}, \citenamefont {Sato},
  \citenamefont {Takahashi}, \citenamefont {Eto}, \citenamefont {Segawa},\ and\
  \citenamefont {Ando}}]{Ando_2014prb}%
  \BibitemOpen
  \bibfield  {author} {\bibinfo {author} {\bibfnamefont {M.}~\bibnamefont
  {Nomura}}, \bibinfo {author} {\bibfnamefont {S.}~\bibnamefont {Souma}},
  \bibinfo {author} {\bibfnamefont {A.}~\bibnamefont {Takayama}}, \bibinfo
  {author} {\bibfnamefont {T.}~\bibnamefont {Sato}}, \bibinfo {author}
  {\bibfnamefont {T.}~\bibnamefont {Takahashi}}, \bibinfo {author}
  {\bibfnamefont {K.}~\bibnamefont {Eto}}, \bibinfo {author} {\bibfnamefont
  {K.}~\bibnamefont {Segawa}}, \ and\ \bibinfo {author} {\bibfnamefont
  {Y.}~\bibnamefont {Ando}},\ }\href {\doibase 10.1103/PhysRevB.89.045134}
  {\bibfield  {journal} {\bibinfo  {journal} {Phys. Rev. B}\ }\textbf {\bibinfo
  {volume} {89}},\ \bibinfo {pages} {045134} (\bibinfo {year}
  {2014})}\BibitemShut {NoStop}%
\bibitem [{\citenamefont {Annese}\ \emph {et~al.}(2018)\citenamefont {Annese},
  \citenamefont {Okuda}, \citenamefont {Schwier}, \citenamefont {Iwasawa},
  \citenamefont {Shimada}, \citenamefont {Natamane}, \citenamefont {Taniguchi},
  \citenamefont {Rusinov}, \citenamefont {Eremeev}, \citenamefont {Kokh},
  \citenamefont {Golyashov}, \citenamefont {Tereshchenko}, \citenamefont
  {Chulkov},\ and\ \citenamefont {Kimura}}]{kimura_2018prb}%
  \BibitemOpen
  \bibfield  {author} {\bibinfo {author} {\bibfnamefont {E.}~\bibnamefont
  {Annese}}, \bibinfo {author} {\bibfnamefont {T.}~\bibnamefont {Okuda}},
  \bibinfo {author} {\bibfnamefont {E.~F.}\ \bibnamefont {Schwier}}, \bibinfo
  {author} {\bibfnamefont {H.}~\bibnamefont {Iwasawa}}, \bibinfo {author}
  {\bibfnamefont {K.}~\bibnamefont {Shimada}}, \bibinfo {author} {\bibfnamefont
  {M.}~\bibnamefont {Natamane}}, \bibinfo {author} {\bibfnamefont
  {M.}~\bibnamefont {Taniguchi}}, \bibinfo {author} {\bibfnamefont {I.~P.}\
  \bibnamefont {Rusinov}}, \bibinfo {author} {\bibfnamefont {S.~V.}\
  \bibnamefont {Eremeev}}, \bibinfo {author} {\bibfnamefont {K.~A.}\
  \bibnamefont {Kokh}}, \bibinfo {author} {\bibfnamefont {V.~A.}\ \bibnamefont
  {Golyashov}}, \bibinfo {author} {\bibfnamefont {O.~E.}\ \bibnamefont
  {Tereshchenko}}, \bibinfo {author} {\bibfnamefont {E.~V.}\ \bibnamefont
  {Chulkov}}, \ and\ \bibinfo {author} {\bibfnamefont {A.}~\bibnamefont
  {Kimura}},\ }\href {\doibase 10.1103/PhysRevB.97.205113} {\bibfield
  {journal} {\bibinfo  {journal} {Phys. Rev. B}\ }\textbf {\bibinfo {volume}
  {97}},\ \bibinfo {pages} {205113} (\bibinfo {year} {2018})}\BibitemShut
  {NoStop}%
\end{thebibliography}%

\end{document}